\pdfoutput=1
\documentclass[11pt,final]{article}
\usepackage{geometry}
\usepackage{showlabels}
\usepackage[utf8x]{inputenc}
\usepackage[T1]{fontenc}
\usepackage{booktabs} %
\usepackage[open=true]{bookmark}
\usepackage{authblk} %

\makeatletter
\def\blfootnote{\xdef\@thefnmark{}\@footnotetext}
\makeatother

\input{macros}
\title{Towards Data Auctions with Externalities}
\hypersetup{colorlinks=true,citecolor=blue}
\date{}

\author[a]{Anish Agarwal}
\author[a]{Munther Dahleh}
\author[a]{Thibaut Horel}
\author[a]{Maryann Rui}
\affil[a]{
Institute for Data, Systems, and Society, Massachusetts Institute of Technology, Cambridge, MA 02139
\protect\\ anish.agarwal@gmail.com,\{dahleh,thibauth,mrui\}@mit.edu}

\begin{document}

\maketitle

\begin{abstract}
The design of data markets has gained importance as firms increasingly use machine learning models fueled by externally acquired training data. A key consideration is the externalities firms face when data, though inherently freely replicable, is allocated to competing firms. In this setting, we demonstrate that a data seller’s optimal revenue increases as firms can pay to prevent allocations to others. To do so, we first reduce the combinatorial problem of allocating and pricing multiple datasets to the auction of a single digital good by modeling utility for data through the increase in prediction accuracy it provides. We then derive welfare and revenue maximizing mechanisms, highlighting how the form of firms’ private information – whether the externalities one exerts on others is known, or vice-versa – affects the resulting structures. In all cases, under appropriate assumptions, the optimal allocation rule is a single threshold per firm, where either all data is allocated or none is.

\end{abstract}
\paragraph{Keywords:} Data Markets, Mechanism Design, Externalities
\paragraph{Declaration of interests:} None
\blfootnote{\copyright\ 2024. This manuscript version is made available under the CC-BY-NC-ND 4.0 license: \url{https://creativecommons.org/licenses/by-nc-nd/4.0/}}

\section{Introduction}\label{sec:intro}
There are two key trends in the current era of statistical analysis: 
(i) an increase in the ease and scale of data collection and exchange due to the growing digitization of modern life;
(ii) an increase in the richness of what one can learn from data via large scale statistical analysis due to improved computing capabilities. 
This is why more and more firms are deploying statistical models to improve their operations and make better decisions regarding, for example, what, when and how much to produce and at what price.

The key differentiating factor that determines the accuracy of these statistical models is access to high-quality data used to train these models, which we henceforth call training data.
However, obtaining relevant training data can be challenging for firms, especially if firms do not have the requisite data infrastructure setup to collect data. Addressing this need, many data services now specialize in collecting, aggregating and cleaning data from a wide variety of external sources, and sell it to firms as a product.

In light of the increased usefulness of training data and  the  growth  of  data  services, firms are increasingly becoming sensitive to the externalities they could face when data is allocated to competing firms.
In this paper, we focus our attention on the setting of negative externalities, though under certain conditions, firms may also experience positive externalities from the sharing of data \citep{raith96}. 
Negative externalities may arise because firms that buy similar data often interact in a downstream market, and the use of data confers on a given firm a competitive advantage which hinders its competitors' performance. 
For example, the data being sold could provide information about consumer preferences in the downstream market, which can help a firm capture a larger share of the market via product differentiation.
Thus, although data is a freely replicable digital good and in principle can be sold to all firms interested in buying it, a data seller aiming to maximize revenue or welfare must consider the competition structure between buyers in deciding which firms to sell data to, how much to sell, and at what price. 
Notably, the  formal  literature  which analyzes mechanisms to exchange training data between a collection of buyers and sellers is sparse. 
Thus, in this paper, we seek to provide a formal answer to the question: 

\vspace{5pt}
\begin{center}
\textit{How does the presence of externalities affect the optimal design of data markets? 
}
\end{center}
\vspace{5pt}

\looseness=-1
This question can be approached by solving two coupled problems:
First, can we define a tractable yet expressive model for how a buying firm values a collection of datasets and for the externality such a firm experiences when a competing firm is allocated data?
Second, given such a model, can we derive efficient or optimal mechanisms to sell data, and how is such a mechanism affected by the presence of externalities?\footnote{We note that many data marketplaces do already exist (e.g., Xignite for financial data, Terbine for IoT sensor data \citep{azcoitia22}).
To the best of our knowledge, these real-world marketplaces tend to offer menu or subscription-based pricing options, but we suspect their pricing strategies are not optimized for welfare or revenue generation given the private valuations of data buyers and possible externalities between them due to downstream competition.
Thus one driver of our approach is to explore contexts in which auction-based mechanisms for selling datasets can be optimal.}
Below, we give an overview of how we tackle these two closely related problems and the answers they provide to the key question posed above. 

\paragraph{How to Model a Buyer's Valuation for Data?}
The challenge of modeling a buyer's utility for data stems from certain characteristics that are intrinsic to virtually all training data used to fit a statistical model:
(i) datasets are freely replicable and so have no inherent scarcity;
(ii) data is not fungible, and in fact its value is intrinsically combinatorial, i.e., different datasets (or different training features for a statistical model) are bound to have correlations in signal leading to sub-additive or super-additive valuations.
In addition, in the presence of externalities, the value a buyer gets from a collection of datasets depends on the data allocated to other competing buyers.
Thus a naive parameterization of a buyer's valuation for a collection of datasets will be exponential in the number of datasets available and in the number of competing firms, rendering such a model intractable.

To circumvent these difficulties,
we recall that a major motivation for firms to buy data is to make better predictions. 
Given this lens and building on the model in \cite{ADS19}, a key assumption we make in our formulation is that {\em a buyer's value for a collection of datasets comes directly from the increase in prediction accuracy it brings to a statistical model}. 
Specifically, we assume that, for each buyer with a given prediction task, there is a commonly known function which maps each collection of datasets to a real number between $0$ and $1$.
This number is a scalar summary of the increase in \emph{prediction accuracy} of the statistical model -- for example, this could be based on the (normalized) mean-squared error of a regression model trained using these datasets.
A buyer's value for a collection of datasets is then a non-decreasing function of this scalar summary of prediction accuracy. 
This formulation is not only crucial for a tractable model of a buyer's valuation over datasets, but also, we believe, a better abstraction for reasoning about the value of data.
The inferences drawn from a dataset, rather than the raw data itself, are what is of interest to the buyer of a dataset.
Indeed, one would typically be willing to trade one dataset for another as long as the same inferences can be drawn from either of the two datasets.

Further, given that our aim is to understand the effect of externalities in a data market, our model needs to incorporate that a data buyer not only has a positive marginal value for acquiring data, but also has a negative marginal value for its competitors acquiring data.
Thus a crucial extension to the model in \cite{ADS19} is that we also assume that the prediction accuracy achieved by a given buyer induces a linear externality on each of the other buyers.
This shift in perspective, from raw data to prediction accuracy, along with assuming a linear externality model reduces the combinatorial valuation of each buyer of data to $n$ parameters (where $n$ is the number of buyers): 
the positive marginal value for the buyer's own prediction accuracy and the negative marginal values for the prediction accuracy of the remaining $(n-1)$ buyers.

\paragraph{The Design and Properties of a Data Market with Externalities.}
Given the data valuation and externality model we introduce, we design a data auction for the setting of a monopolistic data seller and $n$ potential buyers (or bidders).
Since we assume a buyer's valuation for data is captured by the prediction accuracy, we show that the auction design problem reduces to one of selling a single divisible and freely replicable good. 
That is, instead of considering the seller as directly allocating data to buyers, we can equivalently consider the seller as allocating prediction accuracies resulting from the same data allocation.
Still, a buyer's bid is multidimensional since it includes private \emph{externality parameters} in addition to a buyer's marginal value for their prediction accuracy, and this multidimensionality makes characterizing the welfare-maximizing (i.e., efficient) and revenue-maximizing (i.e., optimal) mechanisms particularly challenging. 
Further, in such a market, two natural information structures are possible. 
In the first one, each buyer knows the externalities exerted on them by other buyers, i.e., the buyer's marginal values for other buyers' increases in prediction accuracy. In the latter the buyer knows the externalities that they exert on other buyers. %
We are thus tasked with describing different mechanisms corresponding to the specific information structure and objective being maximized (social welfare or revenue).%

Optimal mechanism design with multidimensional bids is notoriously hard and our setting is no exception. When buyers privately know the externalities exerted on them by other buyers, we prove a reduction between the problem of auctioning multiple items to a single additive buyer and our setting. Understanding the structure of optimal mechanisms for the former problem is a major open question in auction theory (even for just two items and i.i.d.\ valuations). We thus make the problem tractable in this case by restricting either the mechanism structure, or the type distribution. In contrast, when buyers privately know the externalities they exert on other buyers, the problem simplifies and no such restriction is required: we show that the seller ignores the reported externalities and replaces them with their expected values under the common prior.

In all cases, the efficient and optimal mechanisms have a pleasantly simple structure: the allocation rules can be described by thresholds, resulting in allocations, to each given bidder, of either the entirety of the good (equivalently, all the data) or none of it. 
These thresholds balance a buyer's marginal value for their own accuracy with others' (negative) marginal values for the buyer's accuracy. 
We also show that under appropriate assumptions, revenue maximization reduces to \emph{virtual surplus maximization}: the optimal allocation is obtained by replacing the buyers' marginal values with the corresponding virtual marginal values in the efficient allocation. This property was already observed in the context of single-item auctions \citep{M81}, hence our results extend it to the setting with externalities.

The thresholding structure of the optimal allocation has two important robustness implications:
(i) the data seller does not require knowledge of the specific statistical models used by the buyers, nor of each buyer's mapping from data to prediction accuracy, as long as the map is monotonic in the data allocated; 
(ii) if the prior distribution of private types is unknown, the optimal thresholds can be learned via an online optimization framework.
We present such a framework in Appendix \ref{sec:learning} for the case where
the buyers' bids are of the externalities they exert on others. %

We now highlight some key properties of the optimal mechanisms that we derive, and how they are affected by externalities.
First, we establish that in the presence of externalities, the maximum revenue a data seller can collect increases, even if the overall allocation decreases. 
This increase in revenue occurs as the data seller collects payments from firms to prevent allocations to other competing firms.
Second, a perhaps counter-intuitive property of the optimal mechanisms is that the seller will in some cases collect a payment, which can also be viewed as an entry fee, even from buyers to whom nothing is allocated. 
This stems from the implicit threat induced by the allocation to other buyers resulting in a negative utility even when a buyer does not participate in the auction. 
Finally, the payment rules in the presence of externalities extend the intuition of standard second-price auctions. 
Here, bidders will be charged the minimum externality bid they would have needed to report in order to prevent another bidder from receiving an allocation, in addition to the minimum threshold they needed to bid in order to receive their own allocation.
These key properties laid out above provide a meaningful answer to the main question of this paper posed earlier, of how externalities affect the optimal design of data markets.

As our key technical contribution, we extend the Myersonian auction format to the setting of a non-rival, excludable good with multidimensional bids that capture negative externalities.
This provides novel results on the design and properties of the optimal mechanism for such settings.

\paragraph{Outline of the Paper.} 
We conclude the introduction with a discussion
of the related work.  
In Section~\ref{sec:model}, we formulate the utility
model of the firms, discuss our modeling assumptions and present the auction
design problem. In Section~\ref{sec:SW}  and Section~\ref{sec:RM} we describe the welfare-
and revenue- maximizing mechanisms respectively. We conclude with a discussion of our main results in Section~\ref{sec:summary}.
In Appendix~\ref{sec:learning}, we extend our mechanism to when the prior distribution of types is unknown. The remaining appendices contain proofs of our results.

\subsection{Related Work}\label{sec:rel_work}

\paragraph{Economics of Information Goods.} The question of \emph{information
sale} has been the subject of a long line of work going back at least to the
papers of \cite{admatiMonopolistic1986,admatipfleiderer88,admatiDirect1990} and
more recently \cite{kastlSelling2018,BA12,BBS18,BCT19}. In this line of work,
a (most often) monopolist information seller chooses an information
structure—that is, sets of signals and distributions over signals—to offer to
one or more agents engaged in a downstream game of incomplete information. The
information received by the agents allow them to take better actions in the
downstream game, thus increasing their utility. Similar to our results, some of
these works find that under certain regimes of competition between firms, it is
indeed optimal to sell to a strict subset of firms, balancing gains from
information and negative externalities from competition. Importantly, in this
line of work, the agents have no private information, hence the information
seller does not face a screening problem and can extract the entirety of the
surplus generated by their chosen information structure. Consequently, the
seller's problem becomes the one of finding an information structure that
maximizes this surplus.

\cite{raith96,ziv93} considered the related problem of \emph{sharing}
market-relevant information among competing oligopolists, and showed that the
effect of such information sharing on the overall welfare of the firms depends
on the type of competition in which they are engaged (e.g.\ Bertrand or Cournot
competition), and the type of market-relevant parameters they are sharing
(e.g.\ firms' individual production cost estimates or a common market demand
parameter). In some cases it is not optimal for any firm to share information
with the others, due to the overwhelming negative effects of increased
competition on their downstream profit. These findings motivate the study of
how different forms of interdependent valuation functions may affect the
welfare-maximizing or revenue-maximizing allocation of data. 
\cite{hellwigKnowing2009,myattEndogenous2012,amirEndogenous2016} study the
closely related problem of \emph{information acquisition}: there, the
information structure is not designed to optimize a specific objective, but
rather the goal is to understand which signal(s), among a pre-existing
collection, the agents will choose to acquire and analyze the impact of these
signals on the downstream equilibrium.

A related but more recent line of work is the one of \emph{information design}
\citep{elliottMarket2021,armstrongConsumer2022} and the closely
related problem of Bayesian persuasion. There again, an information designer
chooses an information structure to offer to one or more agents engaged in a
downstream game of incomplete information. The main difference with information
sale is one of perspective: the goal is not as much to find the information
structure that maximizes the information provider's profit, but rather to
characterize the set of equilibria that can be induced by the information
structure. Typically, the agents are modeled as firms engaged in a downstream
market, and one wishes to characterize the range of possible market outcomes
(firm-favored vs.\ consumer-favored). As in the case of information sale, the
agents have no private information (prior to receiving a signal from the
information provider) and the problem is purely one of designing the
information structure, without screening.

\looseness=-1
In reality, data buyers may have private informational priors and valuations on
dataset allocations, which calls for the integration of an auction framework
that incentivizes participation and truthful bidding by the buyers. A growing
collection of recent papers
\citep{bonattiCoordination2023,candoganOptimal2023,BDHN21,rodriguezoliveraStrategic2024,segura-rodriguezSelling2022}\footnote{This
line of work postdates the first online appearance of the present paper which 
directly influenced \citep{BDHN21,bonattiCoordination2023}.} extends standard
information design to situations in which the agents have private information
(for example, their cost of taking an action in the downstream game). The
problem becomes one of jointly designing the information structure (often to maximize
a specific objective such as welfare or revenue) while at the same type
eliciting the agents' private information via a truthful direct revelation
mechanism. This is a challenging problem that has so far only been considered
for single-dimensional types, with the exception of
\cite{segura-rodriguezSelling2022} (but where there is only a single buyer).

In summary, compared to these papers, our model and assumptions are less
sophisticated from an information design perspective but richer from a
mechanism design perspective. Indeed, we abstract away the details of
downstream game and hence do not have to reason about the impact of the
allocated information on the downstream equilibrium.\footnote{At a formal
level, this means that we do not have to consider the obedience constraint from
information design.} On the other hand, we explicitly model allocative
externalities and allow agents to express preferences about those. As we will
show, these reduces our problem to the one of allocating a digital (freely
replicable) good to multiple agents in the presence of allocative
externalities. We thus face a multidimensional screening problem which is
absent from the literature on information design with elicitation. This makes
our work closer at a technical level to the literature on auctions with
externalities which we discuss below.

\paragraph{Sale of Data.}Still within the mechanism design literature, but moving away from
information design, another line of work studies mechanism design for the sale
of data, in which the value of data is derived from its informativeness in a
learning task. For procurement auctions, \cite{GR11} consider a setting in
which the buyer wishes to estimate a population statistic while the sellers
experience a cost due to privacy loss. In \cite{RS12}, the authors consider a
similar problem but assume a known prior on the sellers' costs. A
budget-feasible regression problem is considered in \cite{HIM14} and
\cite{ACHW15} consider an online learning setting. \cite{ADS19}  develops a
two-sided market for selling and buying data, capturing the value of data
through increases in prediction accuracy for buyer-specific machine learning
models. In our work, we build on this model of valuation and study auctions of
data in the presence of externalities.

Other recent works look specifically at the sale of consumer data to firms. \cite{BB19} provide an excellent survey by considering a general data market framework featuring data buyers, data sellers, and data intermediaries. Key questions at hand include how data should be priced and possibly versioned, either directly by the data sellers or by the data intermediaries. \cite{TW14} study specific settings in which firms in an oligopoly may use consumer data to set personalized prices, and find that the outcomes of data policies strongly depend on the oligopoly structure. \cite{AMMO19} study a form of externalities between data \textit{sellers} who value their privacy. In their model, correlations between consumer signals yield equilibria where consumers sell their data for very cheap prices despite having high values for privacy.
\cite{BBG20} similarly consider correlations between consumers' data in the context of personalized pricing. The authors find that in order to maximize profits, a data intermediary should only sell an aggregate statistic, rather than individual values, of consumer demand to price-setting firms. 
Like \cite{TW14}, we consider externalities among data \textit{buyers}, but assume a general model of additive negative externalities among the buyers that doesn't necessarily have to arise from competition in oligopoly. %
Finally, \cite{AcquistiPrivacy} provide a comprehensive review on the economic and privacy implications of collecting, using, and selling consumer data from both theoretical and empirical perspectives.

\paragraph{Auctions with Externalities.}
Both efficient and optimal auctions of a single nondivisible good to multiple
buyers are well understood in the absence of externalities among the buyers.
A welfare-maximizing auction is given by the celebrated Vickrey auction
\citep{vickrey61} which is a second-price sealed bid auction, and
a Bayesian-optimal auction for buyers with identical type distributions is
given by a second-price auction with a reservation price \citep{M81}.

The most relevant line of work in the auction literature studies the question of designing auctions in the presence of externalities. The survey \cite{JM06} provides a useful reference.
Optimal single-item auctions with additive allocative externalities among
bidders were studied in \cite{JMS96,JMS99}. They consider the same
multidimensional, interdependent valuation setting as the one presented here,
and provide characterizations of truthful and
individually rational mechanisms. However, in order to solve revenue maximization under the information structure setting of known incoming externalities (defined in Section \ref{sec:firm_private_type}), \cite{JMS99} impose restrictive symmetry assumptions which effectively reduce the problem to one with a single-dimensional bidding structure. In this paper, we provide a hardness result for the general revenue maximization problem and provide two additional conditions which each yield sensible, yet tractable solutions.

Many papers consider a similar additive model, but often assume that
externality parameters are public \citep{AC08,B13} or do not depend on the identity of the competitor
{\citep{BDP17}},
effectively reducing the auction to the single dimensional setting. Closest to our work is \cite{DP11} which extended the setting of \cite{JMS96} to the situation where $K$ copies of the same indivisible item are being sold. 
 However, their focus was on quantifying the effect of changing the parameter $K$. Finally, \cite{HIMM13,ZWWB18}  consider single-dimensional non-additive models of externalities yielding tractable auctions.

\section{Model}\label{sec:model}
\subsection{Data Valuation Model}
 
We consider $n$ firms, in the set $N\eqdef [n]$, acquiring data from a
monopolistic data seller with the goal of increasing the prediction accuracy of
a statistical model. The data owned by the seller is a finite data set $S$, any
subset of which could be sold to any of the buyers. For example, $S$ could be a
collection of training features or database records. Importantly, data is
freely replicable and the seller can for example decide to sell the same subset
$S'\subseteq S$ to two different buyers.

Formally, each firm $i \in N$ is parameterized by a ``Gain Function'', $G_i:
2^S \to [0, 1]$, which is a set function mapping a subset $S'\subseteq S$, to a
measure of the prediction accuracy $G_i(S')\in[0,1]$ (e.g., (normalized) mean
squared error, $R^2$-accuracy) resulting from the acquisition of $S'$. Implicit
in $G_i$ are the particulars of the ML model that is trained and used to make
predictions through the acquired information. A different $G_i$ for each buyer
indicates that we allow each buyer to have a different prediction task or a
different prediction procedure. We impose two natural properties on the gain
function $G_i$ for each firm $i\in N$.
\begin{property}[Monotonicity]\label{property:gain_monotone}
For any two subsets $S_1\subseteq S_2 \subseteq S$, 
$G_i(S_1) \le G_i(S_2)$.
\end{property}

\begin{property}[Normalization]\label{property:gain_normalized}
We assume that $G_i$ is normalized such that $G_i(S) = 1$ and $G_i(\emptyset) = 0$. 
Here $\emptyset$ denotes the empty set, i.e., no information is allocated to firm i.
\end{property}

\noindent
Property~\ref{property:gain_monotone} makes the natural assumption that the gain function $G_i$ is monotone in the amount of information allocated.
Property~\ref{property:gain_normalized} is simply to normalize the various gain functions, and is without loss of generality. 

Given a subset $S_i\subseteq S$ of data allocated to firm $i$, define the allocation variable
$x_i \eqdef G_i(S_i)$ to be the resulting gain in prediction accuracy. Because of
Property~\ref{property:gain_normalized}, we have $x_i\in[0,1]$ where $x_i
= 1$ denotes firm $i$ getting allocated the entire set $S$, and
$x_i = 0$ denotes firm $i$ getting allocated no information.

For each $i \in [N]$, we assume that firm $i$'s valuation for a given vector of
allocations $\bx\eqdef (x_1,\dots, x_n)\in[0,1]^n$ is given by
\begin{align}\label{eq:firm_valuation_model}
\nu_i(\bx) = v_i x_i - \sum_{j \in N \sm i} \eta_{i\gets j} x_j
\,,
\end{align}
where $v_i \in \Rb_{+}$ is firm $i$'s marginal value for an increase in its own
prediction accuracy and the externality parameters $\eta_{i \gets j} \in
\Rb_{+}$ are the marginal values for a decrease in the prediction accuracies of
firms $j\in N\sm\set{i}$.\footnote{We could also allow $\eta_{i \gets j} < 0$
to capture positive externalities. Characterizations of incentive compatible
and individually rational (IR) mechanisms presented in this paper would still
hold, but with a different ``critical type'' defined for IR payments. We focus
on negative externalities in this paper, which already yields a rich set of
phenomena.} Thus, a firm's valuation for data is linear in the gain in
prediction accuracy obtained, and linear in the increases in prediction
accuracies of competing firms, in the form of additive negative externalities.
As we show in Section \ref{sec:examples}.
Equation~\eqref{eq:firm_valuation_model} arises from two natural models of the
downstream competition between firms.

\paragraph{Discussion of modeling assumptions.} 
Inherent in the design of data auctions is the combinatorial nature of data.
Since any subset $S'\subseteq S$ can be selected as a product to be sold, the
number of parameters required to capture the valuation function of each firm
(including the externalities due to other firms) is in general going to be
exponential in the size of the data set $S$ and in the number of firms, likely
rendering this auction intractable.

Following \cite{ADS19}, we circumvent this difficulty by noting that in the
setting where firms aim to increase prediction accuracy, it is natural to make
the modeling assumption that their valuation is not tied to the specific data
that was acquired, but rather from an increase in prediction accuracy of a
quantity of interest.
Indeed, in some settings, firms may have internal knowledge of how to map an increase in their models' accuracy to an increase in revenue. For example, streaming platforms may know that an $x\%$ increase in their ability to predict how likely it is for a given user to watch a given video, translates into a $\$y$ increase in advertising revenue.
While this assumption abstracts away the details of the competition that the firms engage in and bypasses the need to explicitly model the actions taken by the firms as a result of acquiring information, it is not mutually exclusive with an explicit model of the downstream competition (c.f.\ examples in Sections \ref{ex:quality} and \ref{ex:cournot} below).

Further, the assumption of Property~\ref{property:gain_monotone} excludes
situations in which more information hurts—such as when a more informative
experiment (in the sense of mutual information) is worse in the Blackwell
order. However, this is a natural assumption in the setting of firms acquiring
data for the purpose of increasing the accuracy of a statistical model. Indeed,
virtually all estimators studied in statistics come with the guarantee that
their statistical error decreases with the number of samples (e.g., the
standard least squares estimator for linear regression).

Additionally, in the present model, the externality that each firm $i$ exerts on another firm $j$ depends only on a parameter $\eta_{j\gets i}$ and the allocation $x_i$, which is in direct correspondence with the value that firm $i$ receives from its allocation. This allows us to capture how a firm's increase in performance after acquiring data decreases the utility of competing firms, though it precludes settings where a firm's utility may depend on additional factors or nonlinear effects such as the correlation between firms' predictions. 

Finally, note that interactions among data buyers do not always feature negative externalities. For instance sharing fraud data between banks or patient data between hospitals may benefit all parties involved (e.g.,~\cite{rasouli21}), and the trade-off is in the cost of acquiring and sharing data rather than in the negative externalities between buyers.

\subsection{Examples of Downstream Competition}\label{sec:examples}
\subsubsection{Quality-Based Competition}\label{ex:quality}

\looseness=-1
Consider a setting where two firms, indexed by $i \in \set{1, 2}$, such as two smart phone companies, sell similar goods to a population of consumers. Each consumer has a type $\alpha$ representing their sensitivity to a good's quality and they derive a utility $v_i(\alpha) = \alpha \beta_i q_i - p_i $ for buying firm $i$'s good. Here, $q_i$ is the product quality, $p_i$ is the price, and $\beta_i$ is a parameter representing the public perception (e.g., due to advertising effort) of each firm $i \in \set{1, 2}$. Thus, those consumers with higher type $\alpha$ have a higher willingness to pay for an increase in quality. This parameterization of consumer utility has been widely studied in the economic literature \citep{leland77,wauthy96,chambers06}. Note that quality $q_i$ can also be interpreted as the alignment of the design parameters of a firm's good with consumer preferences; it essentially measures how much utility a consumer derives from buying the good. 

Suppose the market is fully covered and each consumer only wishes to purchase one good and selects the firm $i$ that maximizes its utility $v_i(\alpha)$. Then each firm $i$'s net utility, given the price $p_i$, unit production cost $c_i$, and quantity $g_i$ of goods sold is $u_i = (p_i - c_i)\mathbb{P}\set{\alpha: v_i(\alpha) > v_j (\alpha)}$, where the last term represents the fraction of the consumer population that purchases from firm $i$.

Suppose now each firm may buy data to improve the quality of its good, or
equivalently, improve the alignment of design parameters with consumer
preferences, e.g., with new smart phone features, thus increasing $q_i$ to $q_i
+ x_i$. The following result (proved in \cref{appendix:ex_quality}) shows that, under the consumer type distribution studied in \cite{chambers06}, the resultant increase in firms' utilities $\Delta u_i = u_i(q_i + x_i) - u_i(q_i)$ follows the linear model~\eqref{eq:firm_valuation_model}. 

\begin{fact}\label{fact:ex_quality}
Given the quality-based consumer model above where consumer types $\alpha$ are
distributed according to the density function $f(\alpha)=
\ubar{\alpha}/\alpha^2$ for $\alpha \in [\ubar{\alpha}, \infty)$ for some
$\ubar{\alpha} > 0$. Assume that the prices are %
fixed
and that $(p_1-p_2)/(\beta_1 q_1 - \beta_2 q_2)\geq\ubar{\alpha}$. Then, firm
$i$'s valuations $\nu_i(x) \eqdef \Delta u_i$ of an allocation of data $(x_1,
x_2)$ is given by~\eqref{eq:firm_valuation_model} with 
\begin{align*}
	v_i \eqdef \ubar{\alpha}(p_i - c_i) \beta_i/|p_1 - p_2| \text{ and } \eta_{i
	\gets j} \eqdef \ubar{\alpha}(p_i - c_i)\beta_j/ |p_1 - p_2|
\end{align*}
for $i, j \in \set{1,2}$ with $i \neq j$.
\end{fact}

\subsubsection{Cournot Competition}\label{ex:cournot}

In the previous section, we have seen that quality-based competition can be captured exactly by the linear externality model~\eqref{eq:firm_valuation_model}. However, note that such a linear model could also be taken as a first-order approximation of a general valuation function $\nu_i$ with respect to $\bx$, the allocation vector. We now illustrate this in a case of Cournot competition where externalities turn out to be negative (see also~\cite[Section I]{JMS96}). 

Consider two firms, indexed by $i \in \set{1, 2}$, producing perfect substitute goods. Each firm $i$ decides on a production input quantity $q_i \in [0, \infty)$, and has a unit production cost is $c_i \in [0, \infty)$, such that the production cost incurred by firm $i$ is $c_i q_i$. Meanwhile, firm $i$'s production output, or yield, is given by $\alpha_i q_i$, where $\alpha_i \in [0, \infty)$ captures the production efficiency of firm $i$. Let $M \in \R_{+}$ be the market demand parameter, such that the market clearing price is $M - (\alpha_1\cdot q_1 + \alpha_2\cdot q_2)$.
A standard computation gives the equilibrium profits of each firm for a given set of parameters $(\alpha_1, \alpha_2, c_1, c_2, M)$. 

Let us further assume that firms can purchase data from a third party, and that
using such data results in an increase of production efficiency. For
$i\in\{1,2\}$, let $\alpha_i'$ denote firm $i$'s production efficiency after
acquiring the data, the allocation of which is represented by $x_i =
\alpha_i'-\alpha_i$.\footnote{Note, $x_i$ can easily be normalized to lie in
$[0, 1]$ so that Property~\ref{property:gain_normalized} holds. Further, such
an $x_i$ falls within the framework of data markets; our formulation does not
actually require that the gain function is a prediction task, rather a function
that maps allocations of data to utility.} Each firm's valuation for an
allocation of data $(x_1, x_2)$ is the difference in its equilibrium profit
before and after the changes in production efficiency. The next result shows
that to first order, each firm's valuation for $(x_1, x_2)$ follows the linear
externalities model~\eqref{eq:firm_valuation_model}. In particular, the signs
of the coefficients of $x_i$ properly capture the positive value of one's own
data and the negative effects of competition with other firms. 

\begin{fact}\label{fact:ex_cournot}
	\looseness=-1
In the above Cournot model of competition, with an allocation of data
represented by its resultant increase in production efficiency $x_i = \alpha_i'
- \alpha_i$, for $i\in\set{1, 2}$, to the first order in $(x_1, x_2)$, the change in equilibrium profit of firm 1 satisfies \eqref{eq:firm_valuation_model} with nonnegative coefficients
	\begin{equation}
	\label{eq:firm_params}
	\quad v_1\eqdef 4c_1q_1^*/(3\alpha_1)
\quad\text{and}\quad
\eta_{1\gets 2}\eqdef 2\alpha_1 c_2q_1^*/(3\alpha_2^2)\,,
\end{equation}
where $q_1^* = (\alpha _{1}c_{2}-2\alpha _{2}c_{1}+M\alpha _{1}\alpha
_{2})/(3{\alpha _{1}}^2\alpha _{2})$ is the original equilibrium production quantity of firm 1. The result holds similarly for firm 2, with the indices 1 and 2 swapped. 
\end{fact}
The proof of this fact is provided in \Cref{appendix:ex_cournot}. Although any (differentiable) valuation function admits a linear first-order approximation, Eq.~\eqref{eq:firm_params} allows us to verify that externalities are indeed negative, i.e., that $\eta_{1\gets 2}\geq 0$. Furthermore, knowing the explicit dependency of $\eta_{1\gets 2}$ on the problem's parameters will be useful when discussing the information structure below (cf.\ \Cref{sec:firm_private_type}, and Example~\ref{ex:linear} in particular).

\subsection{Firm Private Type}\label{sec:firm_private_type}

Note from \eqref{eq:firm_valuation_model} that firm $i$'s valuation is a function of $(v_i, \be_{i \gets })$.
The private information a firm has, though, may differ from this set of parameters, depending on the details of the downstream interaction of the firms. We call a firm's private information its ``type''.
We consider two natural scenarios of private types:

\vspace{1mm}
{\bf Scenario 1: Knowledge of Incoming Externalities.} 
Firm $i$'s private type is $(v_i, \be_{i \gets })$, with $\be_{i \gets} \eqdef (\eta_{i \gets j})_{j\in N\sm i}$.  
In this case, firm $i$ knows the externalities that other firms exert on firm $i$, which we refer to as firm $i$'s ``incoming externalities''.

\vspace{1mm}
{\bf Scenario 2: Knowledge of Outgoing Externalities.} 
Firm $i$'s private type is $(v_i, \be_{\gets i})$, with $\be_{\gets i} \eqdef (\eta_{j \gets i})_{j\in N\sm i}$. 
In this case, firm $i$ knows the externalities it exerts on other firms, which we refer to as the firm $i$'s ``outgoing externalities''.

\vspace{2mm}
We find that this difference in what defines the private type of a firm, though
subtle, crucially affects the form of the optimal allocation and payment functions.
\tdx{ examples - private type scenarios in two motivating examples [review 3.10, some edits, shortened]}
\begin{example}\label{ex:linear}
Recall the example of quality-based competition in \Cref{ex:quality}. 
In the case where relative firm reputations are negligible ($\beta_i=1$), but production costs $c_i$ are private knowledge to firm $i$, then firms privately know the incoming externalities caused by allocations to other firms, and we are in the setting of Scenario 1. 
On the other hand, if production costs are common knowledge but advertising efforts $\beta_i$ are private information to firm $i$, then we are in the realm of Scenario 2. 

Next, recall the example of two firms in Cournot competition as in \Cref{ex:cournot}. Using the same notations, suppose each firm privately knows its production cost $c_i$, while it shares a common prior (known to all firms and the data seller) on the distribution of all other firms' production costs. Further, suppose all initial production efficiencies $\alpha_i$ are common knowledge, as well as the initial equilibrium production decisions $q_i^*$---for example, if they were observed in a previous season. Then the parameters $v_i$ and $\be_{\gets i}$, in \eqref{eq:firm_params} are privately known to firm $i$. In other words, this example is exactly captured by Scenario 2 above.
\end{example}
\tdx{ $t_i, \hat{t}_i$ definitions; type space, bid space }
\noindent
\paragraph{Bidder Type Spaces and Bid Spaces.}
Going forward, we use standard auction terminology and refer to firms as
bidders.
We denote bidder $i$'s private type as the vector $t_i \in T_i$, where $T_i$ denotes the type space of bidder $i$. 
In Scenario 1, we have $t_i \eqdef v_i e_i - \sum_{j \in N \sm i} \eta_{i \gets j} e_j \in \Rb^n$, while in Scenario 2, we have $t_i \eqdef  v_i e_i - \sum_{j \in N \sm i} \eta_{j \gets i} e_j \in \Rb^n$, where $(e_i)_{i\in N}$ is the standard basis of $\Rb^n$. We abuse notation and let $t_i$ refer to both kinds of private types as it will be clear from context for the remainder of the paper.
We further assume the type values lie in bounded
ranges: $v_i \in [\ubar{v}_i, \bar{v}_i]$ and $\eta_{ij} \in [\ubar{\eta}_{ij},
\bar{\eta}_{ij}]$ for $i \in N, j \in N\sm i$, and let $\ubar{\be}_{i \gets} \eqdef (\ubar{\eta}_{i \gets j})_{j \in N \sm i}$, and $\bar{\be}_{i \gets} \eqdef (\bar{\eta}_{i \gets j})_{j \in N \sm i}$.
Let $T\eqdef\prod_{i\in N} T_i$. A vector of types from all the bidders is denoted
as $\bt \in T$. We denote $\bt_{-i}$ as the vector of all types other than
bidder $i$.

We assume bidders are rational, utility-maximizing agents. It is possible that participating in the auction, i.e.\ submitting a valid bid, receiving an allocation, and making a payment, may leave bidders worse off than simply not participating. To give bidders the option of non-participation, we define the bid spaces $B_i \eqdef T_i \cup \set{\emptyset}$ and $B \eqdef \Pi_{i \in N} B_i$. Then a bidder can report any type in $T_i$ or choose to not participate in the auction by reporting $\emptyset$.

Throughout, we use the convention that a ``hat'' letter denotes a quantity reported by the bidders, as opposed to the ``true'' realization of the same quantity. For example, $t_i$ denotes the (true) type of bidder $i$ while $\hat t_i$ denotes their bid (i.e.\ reported type). Similarly, $\bt_{-i}$ and $\hat\bt_{-i}$ denote respectively the true types and reported types of all bidders but $i$.
\tdx{ distribution on bidder types (independence assumption)}
\paragraph{Prior Distribution of Bidder Types.}
We make the standard assumption that bidders' private types $t_i$ are drawn independently from commonly known distributions $F_i$ on $T_i$, for $i \in [N]$.
Let %
$F = \prod_{i \in N} F_i$ be the joint distribution function of $\bt$ on $T$. For the individual parameters $v_i$ and $\eta_{i\gets j}$, we denote the corresponding marginal density and distribution functions by $f_{v_i}$,  $f_{\eta_{i \gets j}}$, and $F_{v_i}, F_{\eta_{i \gets j}}$, respectively.

\subsection{Auction Design Setup}\label{sec:auctionsetup}
By the revelation principle \citep{M81}, it suffices to
consider incentive compatible mechanisms where bidders directly bid their type.
The auction design problem consists of designing the following two functions to maximize social welfare or the seller's revenue: an \emph{allocation} function $\bx:B\to[0,1]^n$ and a \emph{payment} function $\bp:B\to(\R_{+})^n$.
In short, given a vector of bids $\hat{\bt}\in T$ from the bidders, $\bx(\hat{\bt})$ is the resulting vector of allocations and $\bp(\hat{\bt})$ is the vector of payments required of the bidders. Abusing notation, we let $\bx$ denote both the vector of allocations and the function, which maps bids to this allocation vector, and similarly for $\bp$.

We assume bidders have \textit{quasilinear} net utility from participating in the auction. That is, given allocation and payment vectors $\bx$ and $\bp$, respectively, and true types $\bt \in T$, bidder $i$'s utility is 
\begin{displaymath}
	u_i(\bx, \bp; \bt)
	\eqdef \nu_i(\bx) - p_i
=v_i \cdot x_i - \sum_{j \in N \sm i} \eta_{i \gets j} x_j - p_i	\,.
\end{displaymath}

\begin{remark}[Mapping allocations to data subsets.]
Since we modeled $G_i:2^S\to [0,1]$ as a set function, it can only take a
discrete set of values in the range $[0,1]$,
leading to possible discontinuities in the allocation.
In our auction setup, we relax the problem and consider allocations in the continuous domain $[0,1]$, i.e.\ we will treat in the analysis the allocation $\bx$ as being able to take \emph{any} value in this domain.
Given an allocation $x_i \in [0,1]$, the reverse problem of determining a subset $S_i \subseteq S$ corresponding to $x_i$ can be addressed in several ways. Since $G_i$ only takes finitely many values in $[0,1]$, there may not always exist a subset $S_i$ such that $x_i = G_i(S_i)$. However, one can construct probabilistic allocations or add noise to the data to interpolate between values. 

As we will prove (c.f.\ Table \ref{table:summary} in Section \ref{sec:summary} for a summary of results), it turns out that even though we relax the problem to the continuous allocation setting, the optimal allocations (for both welfare and revenue maximization), under appropriate assumptions, are single price thresholds (one per firm),
above which the seller allocates all information and below which, allocates no information to a firm. Thus, the mechanism remains realizable for the original problem with discrete allocations and we conveniently avoid the issue of these discontinuities by having to only implement the extremal allocations, $G_i(S)$ (i.e., $x_i = 1$) and $G_i(\emptyset)$ (i.e., $x_i = 0$). 
Our results of all-or-nothing allocations have the additional benefit that the data seller does not need to know the form of $G_i$, just that Property~\ref{property:gain_monotone} holds. 
\end{remark}

\begin{remark}%
The key difference from standard single-item auction setups is that for digital goods, such as data, there is no feasibility constraint on the allocation function
$\bx(\cdot)$.
In particular, we do not require that the sum of the allocations ($\sum_{i = 1}^n x_i$), is less than or equal to one.
The absence of this feasibility constraint is key in 
	obtaining a simple structure for the optimal auctions despite it being
a multi-dimensional mechanism design problem (i.e., each bidder is parameterized
by a $n$-dimensional vector).
\end{remark}
\noindent
\paragraph{Outside Option.}
When a bidder chooses not to participate in the auction, the auctioneer cannot charge the bidder any price nor `dump' any goods on the bidder. That is, we have the restriction that $x_i(\hat{t}) = 0$ and $p_i(\hat{t}) = 0$ whenever $\hat{t}_i = \emptyset$.
Note that even if a given bidder $i$ chooses not to participate in the auction, allocations to the other, participating bidders can still affect bidder $i$'s utility through negative externalities. 
Thus, it will be necessary to specify what the auction does when subsets of
bidders don't participate. 

Since we are interested in finding a Nash equilibrium
in which all bidders participate (and bid truthfully), it suffices for us to
explicitly define the mechanism under single-bidder deviations from equilibrium
and the equilibrium itself. Note we do not consider coalition-proof or strong Nash equilibria, which may not exist. Thus, we seek allocation and payment rules
$\bx(\hat{\bt})$ and $\bp(\hat{\bt})$ when at most one component of $\hat{\bt}$ is
$\emptyset$. 
Bidder $i$'s ``outside option'' denotes the setting where bidder $i$ do not participate and all remaining bidders $N\sm i$ do participate. Bidder $i$'s outside option utility, or reservation utility, depends only on others' bids and the true underlying
types. Specifically, given a type vector
$\bt\in T$ and a vector of bids $\hat\bt_{-i}$ from other bidders, the outside option utility
of bidder $i$ is given by
\begin{displaymath}
u_i\of[\big]{\bx(\hat{t}_i = \emptyset, \hat{\bt}_{-i}), \bp(\hat{t}_i = \emptyset, \hat{\bt}_{-i}); \bt}
= -\sum_{j \in N \sm i} \eta_{i \gets j}
x_j(\hat{t}_i = \emptyset, \hat{\bt}_{-i})
\,.
\end{displaymath}

\subsubsection{Definitions of IC and IR Mechanisms}%
We now define the incentive compatibility (IC) and
individual rationality (IR) constraints that the mechanisms must satisfy.
\looseness=-1

\paragraph{Ex-Post Constraints.} We first consider ex-post
truthfulness and participation constraints.
\begin{definition}[Dominant Strategy Incentive Compatibility] \label{def:dsic}
	A mechanism $(\bx, \bp)$ is \emph{Dominant Strategy Incentive Compatible}
	(DSIC) if for all type vectors $\bt,\hat\bt\in T$ and bidder
	$i\in N$
\begin{displaymath}
u_i\of[\big]{\bx(t_i, \hat\bt_{-i}), p_i(t_i,\hat\bt_{-i}); \bt}
\geq u_i\of[\big]{\bx(\hat\bt), p_i(\hat\bt); \bt}
\,.
\end{displaymath}
\end{definition}
\begin{definition}[Ex-Post Individual Rationality]\label{def:expostIR}
A mechanism $(\bx, \bp)$ is \emph{ex-post Individually Rational} (ex-post IR)
if for every type vector $\bt\in T$ and bidder $i\in N$
\begin{displaymath}
u_i\of[\big]{\bx(\bt), p_i(\bt);\bt}
\geq u_i\of[\big]{\bx(\emptyset, \bt_{-i}), p_i(\emptyset, \bt_{-i});\bt}
\,.
\end{displaymath}
\end{definition}
Dominant strategy incentive compatibility expresses that no matter what the
true types are and what other players bid, a bidder cannot strictly increase
her net utility by bidding untruthfully. Ex-post individual rationality
expresses that no matter what the true types are, in a situation where all
other bidders participate and bid truthfully, it is better for each bidder to
report truthfully than to not participate. These two properties combined imply
that participating and reporting truthfully is a dominant strategy equilibrium
of the game induced by the mechanism.

\paragraph{Interim Constraints.}
In situations where types are drawn from a known prior distribution and bidders
reason in expectation over other bidder's private types conditioned on their
own observed types, we consider \emph{interim} relaxations of the IC and IR
definitions.
To this end, define $V_i(\hat{t}_i; t_i)\eqdef
\cex*{u_i\of[\big]{\bx(\hat{t}_i, \bt_{-i}), p_i(\hat{t}_i, \bt_{-i}};
\bt)}{t_i}$ to be the interim expected utility of bidder $i \in N$ if they bid
$\hat{t}_i \in B_i$ with a true type of $t_i \in T_i$, and all other
bidders bid their type truthfully. Note that the expectation is taken over
a random realization $\bt\sim F$ conditioned on the event that bidder's $i$
type is $t_i$.

\begin{definition}[Bayes--Nash Incentive Compatibility]\label{def:bnic}
A mechanism $(\bx, \bp)$ is \emph{Bayes--Nash Incentive Compatible} (BNIC) if
for all types $t_i,\hat t_i\in T_i$ and bidder  $i\in N$,
$ V_i(t_i; t_i) \geq V_i (\hat{t}_i; t_i)$.
\end{definition}
\begin{definition}[Interim Individual Rationality]\label{def:interimIR}
A mechanism $(\bx, \bp)$ satisfies \emph{interim Individual
Rationality} (interim IR) if for every type $t_i\in T_i$ and bidder $i\in N$,
$ V_i(t_i; t_i) \geq V_i(\emptyset; t_i)$.
\end{definition}

Appendix \ref{sec:IC_IR_char} provides relevant characterizations of the payment and allocations functions induced by the interim IC and IR constraints, which depend on the form of bidders' private types.

\section{Social Welfare Maximization}\label{sec:SW}

In this section, the seller's problem is to design allocation and payment functions,
$\bx(\cdot)$ and $\bp(\cdot)$ in order to maximize the welfare, i.e.\ the
sum of bidder valuations: 
\begin{equation} \label{eq:sw_main}
	\sw(\bx; \bt)
	= \sum_{i \in N} \nu_i(\bx)
	= \sum_{i \in N}\Big(v_i x_i - \sum_{j\in N\sm i}\eta_{i\gets j}x_j\Big)
\end{equation}
such that the auction: (i) is incentive compatible; (ii) satisfies individual
rationality; (iii) has no positive transfers, i.e.\ the seller never pays a
bidder to participate in the auction. We organize this section by the private
types of the bidders according to the two scenarios described in
\Cref{sec:firm_private_type}.

\subsection{Welfare Maximization in Scenario 1: Known Incoming Externalities}\label{sec:sw1}
We first consider the setting where the private type of bidder $i \in N$ takes the form $t_i = v_i e_i + \sum_{j \in N \sm i} \eta_{i \gets j} e_j$, i.e., each bidder $i$ knows the incoming allocative externalities $\be_{i \gets}$ that others exert on bidder $i$. We instantiate the
Vickrey--Clarke--Groves (VCG) mechanism and discuss the
resulting allocation and payment functions.

We wish to maximize~\eqref{eq:sw_main} subject to DSIC (\Cref{def:dsic}), ex-post IR (\Cref{def:expostIR}), and the feasibility constraint that for all $i \in N, x_i \in [0,1]$. To define ex-post IR, recall that we need to instantiate the outside option, i.e.\ what occurs if bidder $i$ chooses not to participate in the auction. Here, we choose the natural outside option, that is to run the welfare-maximizing auction with the remaining set $N\sm i$ of bidders.

\begin{theorem}[Efficient Mechanism, Scenario 1]\label{prop:vcgprops}
	The mechanism with allocation function 
	\begin{align} \label{eq:sw1_x}
	x_i(\bt) = \ind{W_i \geq 0}
	= \ind[\bigg]{v_i - \sum_{j \in N \sm i} \eta_{j\gets i}
		\geq 0}\,,
\end{align}
outside option $x_j(t_i = \emptyset, \bt_{-i}) = \ind*{W_j^i \geq 0}$
 and payment function 
\begin{align} \label{eq:sw1_p}
p_i(\bt) &= \sum_{j \in N \sm i} \left( W_j^i\big[\ind{W_j^i \geq 0} - \ind{W_j\geq 0}\big] + \eta_{j\gets i} \ind{W_i \geq 0} \right),
\end{align}
where $W_j^i$ is defined as the welfare contribution of bidder $j$ when (only) bidder $i$ chooses to not participate in the auction to be, for $j \in N \sm i$:
\begin{displaymath} 
	W_j^i \eqdef v_j - \sum_{\mathclap{k \in N \sm \set{i,j}}} \eta_{k\gets j},
\end{displaymath}
maximizes the welfare among all DSIC and ex-post IR auctions, and has no positive transfer. 
\end{theorem}

A full proof of this theorem is given in \Cref{appendix:vcgprops}. The result is quite intuitive: if a bidder $i$'s contribution to the overall welfare, $W_i$, is positive, then bidder $i$ receives the good. Each bidder's payment can be interpreted as the sum of the change in welfare if they leaves the auction and the sum of externalities they induces in the current allocation.  %

\paragraph{}
A feature of the welfare-maximizing VCG mechanism in our setting is that it does not guarantee that each bidder's net utility will be nonnegative, but rather no less than the bidder's reservation utility, which could be negative due to externalities. 
Further, while we choose the outside option to be the welfare-maximizing auction with the remaining bidders, as is natural, we could instead have declared the ensuing auction to have any feasible allocation rule for the bidders $N\sm i$ that does not depend on bidder $i$'s bid.
For instance, a feasible outside option is to allocate all data to every $j \in N \sm i$ if bidder $i$ does not participate, resulting in utility $u_i(\emptyset, \bt_{-i}; t_i, \bt_{-i}) = -\sum_{j \in N \sm i}\eta_{i\gets j}$.
This is in fact the worst possible outside option for bidder $i$, which thereby increases the set of IR-satisfying mechanisms. Indeed, as discussed in \cref{sec:RM}, this worst-case outside option is the revenue-optimal one.

\subsection{Welfare Maximization in Scenario 2: Known Outgoing Externalities}\label{sec:sw2}

We now consider the case where each bidder $i \in N$ knows the externalities $\be_{\gets i}$ that they would exert on other bidders if allocated the good, i.e., where the private type of bidder $i$, is $t_i =v_i e_i - \sum_{j \in N \sm i} \eta_{j \gets i} e_j$.

Note that in this scenario, bidder $i$ cannot fully evaluate their valuation of a given allocation $\bx$, as it depends on the parameters $\be_{i\gets}$, which are part of the private types of bidders $j\in N\sm i$. Each bidder can only reason with their own realized type $t_i$ and the commonly known priors on other bidders' types. It is more sensible, therefore, to impose interim versions of truthfulness (BNIC) and participation (interim IR) conditions (see \cref{def:bnic,def:interimIR} respectively).

\paragraph{Ex-Ante Welfare Optimality.}
A first attempt toward a welfare-maximizing mechanism here may try to use the VCG allocation rule~\eqref{eq:sw1_x} that maximizes welfare pointwise. Due to \cref{prop:BNIC_char}, however, this allocation violates BNIC when the private types are of the form $t_i =  v_i e_i - \sum_{j \in N \sm i} \eta_{j \gets i} e_j $, since the corresponding interim allocation $y_i(t_i) = \ind{v_i \geq \sum_{j \in N \sm i} \eta_{j\gets i}}$ is not in general constant with respect to $\eta_{j\gets i}$. 
In fact, any attempt to find such welfare-maximizing BNIC mechanisms will fail. In general, no mechanism satisfying BNIC can be ex-post (pointwise) welfare-maximal over all types $\bt$, as stated next. 
\begin{proposition} [Impossibility of Ex-Post Optimality]\label{prop:sw2opt-impossibility}
Suppose private types are of the form $t_i = v_i e_i - \sum_{j \in N \sm i}
\eta_{j \gets i} e_j $ for each bidder $ i \in N$. For any joint distribution
$F$ of types $\bt = (t_1,\dots, t_n)$, let $\mathcal{X}_{BNIC}(F)$ be
the set of allocation functions implementable in Bayes-Nash equilibrium. Then
there exists a distribution $F$ of types on $T$ such that
\begin{align} \label{eq:sw_ineq}
\prn[\big]{\forall \bx \in \mathcal{X}_{BNIC}(F)}\,
\prn[\big]{\exists \bt^0\in T, \bx' \in \mathcal{X}_{BNIC}(F)}\::\:
\sw(\bx; \bt^0) < \sw(\bx'; \bt^0).
\end{align}
\end{proposition}
A proof of this proposition is provided in \Cref{appendix:sw2_impossibility}. In particular, it makes use of the characterization of BNIC mechanisms given in \Cref{prop:BNIC_char} of \Cref{appendix:BNIC_char}.
The core of this incompatibility result relies on the multi-dimensionality of signals capturing the allocative externalities. Intuitively, because the parameter $\eta_{j\gets i}$ in bidder $i$'s report does not directly appear in bidder $i$'s valuation function, the BNIC constraint prevents mechanisms from eliciting the true value of $\eta_{j\gets i}$, thus precluding ex-post efficiency. Such an incompatibility between efficiency and BNIC for multi-dimensional types has been more generally studied in \cite{JM01}. Note that in the simpler case of selling a single good with interdependent valuations captured by \emph{one-dimensional} signals, the generalized VCG mechanism \cite{krishna-auctionth} provides an ex-post IC and efficient allocation. %

Since Proposition~\ref{prop:sw2opt-impossibility} implies that there are distributions in which no mechanism satisfying BNIC can also be welfare-maximizing over all type realizations, we relax the objective of finding a pointwise optimum to one of maximizing the \textit{expected} 
welfare, that is,
\begin{align} \label{eq:sw2_sw}
	\ex*{\sw(\bx; \bt)}
= \sum_{i \in N} \ex[\Bigg]{\bigg(v_i - \sum_{j \in N \sm i} \eta_{j \gets i}\bigg) x_i(\bt)}
\,.
\end{align}
Under this relaxed optimality condition, the following theorem describes the mechanism that maximizes welfare in expectation rather than pointwise. 

Suppose bidders have private types of the form $t_i = v_i e_i - \sum_{j \in N \sm i} \eta_{j \gets i} e_j $ for $i \in N$. Define the virtual valuation functions $ \phi_i(v_i) \eqdef
	v_i - \big(1-F_{v_i}(v_i)\big)/f_{v_i}(v_i) $ for $i \in N$. Suppose the functions $\widetilde\phi_i: v_i\mapsto\phi_i(v_i) -\sum_{j\in N\sm	i}\E[\eta_{j\gets i}\,|\, v_i]$ are non-decreasing, and define the thresholds $\tau_i \eqdef \widetilde\phi_i^{-1}(0)$.

\begin{theorem} [Efficient Mechanism, Scenario 2]\label{prop:sw2_alloc} %
	Suppose that the functions $\widetilde\phi_i: v_i \mapsto v_i - \sum_{j \in N \sm i} \E[\eta_{j \gets i} | v_i]$ are non-decreasing, for $ i \in N$, and define thresholds $\tau_i \eqdef \widetilde\phi_i^{-1}(0)$. 
	Then, the allocation rule maximizing the expected welfare~\eqref{eq:sw2_sw} under the BNIC constraint is
\begin{align} \label{eq:sw2_expostalloc}
x_i(\bt) = \ind[\bigg]{v_i \geq \sum_{j \in N \sm i} \cex{\eta_{j\gets i}}{v_i}} = \ind*{v_i \geq \tau_i}
, \,  i \in N\, ,
\end{align}
where the outside option is set to run the welfare-maximizing allocation on the remaining set of bidders whenever some subset of bidders chooses not to participate in the auction.

A class of BNIC and IR payment rules associated with this allocation is given by
\begin{align}
	\label{eq:sw2_expostp}
	p_i(t_i) &= \tau_i \cdot \ind[]{v_i \geq \tau_i}
- \sum_{j \in N \sm i}\ex[\big]{\eta_{i \gets j}\cdot \ind[]{v_j \geq \tau_j}} -
C_i.
\end{align}
where $C_i$ is a constant satisfying %
$C_i = V_i(t_i; t_i)\geq V_i(\emptyset; t_i)$, for some $t_i$ of the form $(\ubar{v}_i, \be_{\gets i})$. In particular, if $v_j \mapsto v_j - \sum_{k \in N \sm \set{j, i}} \E[\eta_{k\gets j}|v_j]$ is non-decreasing for $j \in N \sm i$, then 
\begin{equation}
	\label{eq:sw2_outside}
V_i(\emptyset; t_i)=
\sum_{j \in N \sm i}\ex[\Bigg]{\eta_{i \gets j}
\cdot\ind[\bigg]{v_j \geq \sum_{\mathclap{k \in N \sm \set{{j,i}}}} \cex{\eta_{k\gets j}}{v_j}}}
\end{equation}
\end{theorem}

The proof of \cref{prop:sw2_alloc} is provided in \cref{appendix:sw2_alloc_proof}.
Similar to the result of Theorem \ref{prop:vcgprops}, the allocation rule can be interpreted as allocating the good (i.e., the entire data set) to bidders $i\in [N]$ who, conditioned on their $v_i$, have an expected positive contribution to the social welfare. As discussed regarding Proposition~\ref{prop:sw2opt-impossibility}, a BNIC mechanism must ignore the reported bids of outgoing externalities, so they appear in expectations. Furthermore, the payment rule can be interpreted as each bidder paying the minimum threshold they needed to bid (in $v_i$) to receive the good, less the expected externalities they suffer from other bidders who are allocated the good, plus a constant that is set by the outside option.

\begin{remark}
Note that if we were instead selling a non-replicable good, the feasibility constraint $\sum_{i \in N} x_i \leq 1$ would
couple the allocations and $x_i$ would be a function of other bids $v_j$ for
$ j \neq i$.
\end{remark}

\section{Revenue Maximization}\label{sec:RM}

In this section, we focus on the problem of designing auctions that achieve
optimal revenue. Specifically, the goal is to design allocation
and payment functions $\bx(\cdot)$ and $\bp(\cdot)$ to maximize the
seller's expected revenue
$
\rev(\bx, \bp) \eqdef \sum_{i \in N} \ex[\big]{p_i(\bt)}
$
subject to BNIC and interim IR constraints.

\newcommand\tv{t^\dagger} %
\subsection{Revenue Maximization in Scenario 1: Known Incoming Externalities}\label{sec:rm1}

To begin, we study the revenue maximization problem in the setting where each bidder knows the externalities they incur from other bidders. Throughout this section, we write the type $t_i$ of bidder $i$ as $t_i = v_i e_i -\sum_{j\in N\sm i}\eta_{i\gets j}e_j\in\R^n$.

\subsubsection{Hardness result} Our first result shows that finding the revenue-optimal mechanism in this setting is generically as hard as finding the optimal mechanism for the auction of multiple items to a single buyer with an additive utility function.
This is a negative result in that the latter problem is notoriously hard, both from a mathematical and computational perspective, even in the simple setting of i.i.d.\ item valuations. \citet[Sec.~3]{daskalakis2015} gives a good exposition of the main obstacles:
\begin{itemize}
	\item Even though the items' valuations are independent, pricing each item separately is not always optimal and it can be necessary for the optimal mechanism to \emph{bundle} a subset of the items together.
	\item The optimal mechanism is not always deterministic and sometimes requires offering \emph{lotteries} over several bundles.
	\item Even when the item distributions are described by a finite number of parameters, the optimal mechanisms can require offering uncountably many lotteries.
\end{itemize}
In light of these obstacles, it is perhaps not a surprise that the optimal
mechanism is known in just a handful of special cases. This hardness also
manifests itself in the algorithmic realm where computing the optimal mechanism
is known to be $\#\text{P}$-hard \citep{daskalakis2014complexity}. The next
proposition shows that all these hardness results extend to the auction of a
digital good with additively separable externalities by establishing a
reduction from the multi-item auction.\footnote{We are deeply indebted to
Haifeng Xu and an anonymous referee for this argument. A previous version of
this paper contained a derivation of the revenue-optimal mechanism that was in
contradiction with \Cref{prop:impossibility} and therefore incorrect. In
\Cref{prop:NEWrm2_BNIC} below, we show that this former mechanism is optimal
among a restricted class of mechanisms.}

\begin{proposition}\label{prop:impossibility}
The problem of finding a revenue-optimal mechanism for the auction of $n$ items to a single additive buyer reduces to optimally selling a (freely replicable) digital good to $n$ bidders with additively separable externalities.
\end{proposition}

The proof of \cref{prop:impossibility} is provided in \cref{app:imposs}. The
main idea is to construct an instance of the $n$-bidder digital good auction
from an instance of the $n$-item auction as follows: bidder $1$'s type
distribution is identical to the buyer in the $n$-item auction, and the type
distribution of the remaining $n-1$ bidders is supported on the zero vector.
In this case, the valuation over allocations of each of the $n-1$ “dummy” bidders is a constant
equal to zero, so no payment is collected from them, and the only thing that
matters is the effect of their allocations on bidder $1$'s utility. The
allocation of item $j$ to the buyer in the $n$-item auction, becomes
equivalent, through the reduction, to the allocation to bidder $j$ in the
$n$-bidder auction.\footnote{Some adjustments are necessary due to the sign
discrepancy between externalities and item valuations, and the negative
reservation utility in the $n$-item auction.} The proof is complete after
checking that the reduction preserves revenue as well as the IC and IR
constraints.

\paragraph{} Given this hardness result, we will not attempt to solve our mechanism problem in the most general case. Instead, we consider two structural assumptions under which we can solve for the optimal mechanism:
\begin{enumerate}
\item \emph{Restricted-dependency mechanisms.} In \cref{sec:restricted}, we limit our search to direct-revelation mechanisms whose allocation to bidder $i$ only depends on the parameters capturing the direct effect of bidder $i$'s allocation on welfare. Formally, we assume that $x_i(\bt) = h_i(v_i, \set{\eta_{j \gets i}}_{j\in N\sm i})$ for some function $h_i$ and each $i\in N$.

This assumption, though strictly suboptimal in general, circumvents computational difficulties that arise from the option of bundling allocations (as in optimal multi-item auctions), and is still meaningful in that it allows the mechanism to incorporate the preferences on the allocation to a given bidder from all affected parties.

\item \emph{Single-dimensional types.} In \cref{sec:scen1-single-dim-type}, we assume each bidder $i'$s type vector is maximally correlated in the sense that $\eta_{i\gets j}= \alpha v_i$ for every $j \in N\sm i$, where $\alpha$ is a publicly known scalar multiplier common to all bidders. In this case, bidder $i$'s type can be parameterized by the single-dimensional value $v_i$. 

\end{enumerate}

\subsubsection{Restricted-dependency mechanisms}\label{sec:restricted}
\newcommand\yi{y^{(i)}}
\newcommand\gi{g^{(i)}}
\newcommand\that{\hat{t}}
\newcommand\vhat{\hat{v}}
We first consider revenue maximization over a simpler class of mechanisms $(x, p)$ for which the allocation to each bidder $i \in N$ is only a function of the parameters that directly capture the effect of bidder $i$'s allocation on welfare. Formally, we assume that $x_i(\bt) = h_i(v_i, \be_{\gets i})$ for some function $h_i$ and each $i\in N$.

\begin{theorem}\label{prop:NEWrm2_BNIC}
Suppose bidders have private types of the form $t_i = v_i e_i - \sum_{j\in N\sm i} \eta_{i \gets j} e_j$ for $i \in N$, and consider the class of restricted-dependency mechanisms, where allocation functions are restricted to the form $x_i(t) = h_i(v_i, \be_{\gets i})$, for some functions $h_i$, $i\in N$.

When the virtual valuation functions $\phi_{v_i}(v) \eqdef v -
(1-F_{v_i}(v))/f_{v_i}(v)$ and $\phi_{\eta_{i \gets j}}(\eta) \eqdef \eta + F_{\eta_{i \gets j}}(\eta)/f_{\eta_{i \gets j}}(\eta)$ are nondecreasing, an allocation rule that maximizes the expected revenue among restricted-dependency BNIC and interim IR mechanisms is given by
\begin{equation}\label{eq:rm1_alloc}
	\begin{aligned}
	& x_i(\bt) = \ind[\bigg]{\phi_{v_i}(v_i) \geq \sum_{j \in N \sm i} \phi_{\eta_{j \gets i}}(\eta_{j \gets i})}, \; \text{for } \bt \in T
 \\ &x_j(t_i = \emptyset, \bt_{-i}) = 1, \; \text{ for } j\neq i\text{ and } \bt_{-i} \in T_{-i}.
\end{aligned}
\end{equation}
\end{theorem}

The proof of \cref{prop:NEWrm2_BNIC} can be found in \cref{app:rm-alloc} and relies on a characterization of BNIC mechanisms tailored to restricted-dependency mechanisms (\cref{lem:restricted-BNIC}). Specifically, under the restricted-dependency assumption and due to the mutual independence of bidders' types, the structure of the interim allocation vector field $\by^{(i)}(t_i) \eqdef \cex{\bx(t_i, \bt_{-i})}{t_i}$ disentangles in such a way that its $j$th component only depends on the $j$th component of the input vector $t_i$, for $j \in N$.
As a result, the characterization of BNIC simplifies to (i) requiring monotonicity of each of the component interim functions $y_j^{(i)}$ for $(i,j)\in N^2$, and (ii) pinning down the derivative of the interim payment. The latter lets us express, up to an additive constant, the expected revenue solely in terms of the allocation:
\begin{displaymath}
	\ex[\Bigg]{\sum_{i \in N} x_i(t)\prn[\bigg]{\phi_{v_i}(v_i)-\sum_{j\in N\sm i}
	\phi_{\eta_{j\gets i}}(\eta_{j\gets i})}}.
\end{displaymath}
From this expression, the form of the optimal allocation follows in a straightforward manner. 

\paragraph{}Observe that the allocation rule given in
\Cref{prop:NEWrm2_BNIC} is similar in form to the threshold functions derived in the two social-welfare maximization cases \labelcref{eq:sw1_x,eq:sw2_expostalloc} but with the virtual values now playing the role of the relevant coordinates of the bidders' private types.
In other words, the revenue-maximizing allocation is the one that maximizes \emph{virtual welfare}.
As is typical in revenue maximization settings, the optimal allocation is not efficient in general, that is, allocates the digital good to bidders less often than the welfare-maximizing allocation.
An illustrative example is discussed in \cref{sec:summary}. 

Next, we present an optimal threshold-based payment function that implements
the allocation of \cref{prop:NEWrm2_BNIC}. 

\newcommand{\tauij}{\tau_{ij}}%
\newcommand{\tauii}{\tau_{ii}}%

\begin{corollary}\label{cor:restricted-dep-cor}
Under the assumptions of \cref{prop:NEWrm2_BNIC}, an optimal payment rule that
implements the allocation \eqref{eq:rm1_alloc} is given by
\begin{equation}\label{eq:bnic1_payment}
	p_i(\bt) = x_i(\bt) \cdot \tauii(\bt_{-i}) + \sum_{j \in N \sm i} \big(1-x_j(\bt)\big) \cdot \tauij(\bt_{-i}), 
\end{equation}
where we define the following thresholds
\begin{displaymath}
\tauii(\bt_{-i}) \eqdef \phi_{v_i}^{-1}\of[\Big]{\sum_{j \in N \sm i} \phi_{\eta_{j \gets i}}(\eta_{j\gets i})}
	\quad\text{and}\quad
\tauij(\bt_{-i}) \eqdef \phi_{\eta_{i \gets j}}^{-1}\of[\Big]{\phi_{v_j}(v_j) -\sum_{\mathclap{k\in N \sm \set{i,j}}} \phi_{\eta_{k \gets j}}(\eta_{k \gets j})}.
\end{displaymath}
\end{corollary}

The proof of \cref{cor:restricted-dep-cor} is given in \cref{proof:cor:restricted-dep-cor}. We start from the formula for the interim payment $q_i(t_i)\eqdef \cex{p_i(t_i, \bt_{-i})}{t_i}$ in our BNIC characterization (\cref{lem:restricted-BNIC}), where it is determined, up to an additive constant, in terms of the interim allocation.
As prescribed by \cref{prop:IRchar_type1}, we optimally set the additive constant by making the participation constraint of bidder $i$ bind at their “critical type” $\circt_i\eqdef\ubar{v}_i e_i - \sum_{j \in N \sm i} \ubar{\eta}_{i \gets j} e_j$.
A valid choice for the payment function is to simply equate it with the interim payment pointwise: $p_i(\bt) = q_i(t_i)$. We choose instead the more interpretable form \eqref{eq:bnic1_payment}, in which the payment is expressed in terms of the ex-post allocation, and show in the proof that it still integrates to the same interim payment. As a result, the mechanism remains BNIC and results in the same optimal revenue.

The latter form of payments \eqref{eq:bnic1_payment} has the following simple interpretation: once we fix the types $\bt_{-i}$ of all the bidders but $i$, the thresholds $\tauii$ and $\tauij$ determine, respectively, the minimum bid of $v_i$ that guarantees allocation of the good to bidder $i$, and the minimum bid of $\eta_{i\gets j}$ that prevents bidder $j$ from being allocated the good. Whenever a coordinate of $t_i$ is high enough to make one these “favorable” events occur, the corresponding threshold bid is added to bidder $i$'s payment. This extends the intuition of second price auctions, wherein bidders pay the minimum bid needed to receive the good, to the current setting with externalities.

\subsubsection{Single-dimensional types}\label{sec:scen1-single-dim-type}

Next, we consider the case where the externality parameters in each bidder's type vector are proportional to the bidder's value for being allocated the good.
That is, we assume that for each bidder $i \in N$ and each $j\in N\sm i$, $\eta_{i\gets j}= \alpha v_i$, where $\alpha$ is a publicly known constant, common to all bidders. The sensitivity of bidder $i$'s utility to allocations to other bidders is directly correlated with how much bidder $i$ values the good for themselves.
Under this assumption, each bidder $i$'s type, $t_i = v_i e_i - \alpha v_i e_j$, is effectively one-dimensional.
We reject any bids $\hat{t}_i$ for which $\hat{\eta}_{i \gets j} \neq \alpha v_i$, so that valid bids are essentially reports of $v_i$, and let $\bv = (v_1, \dots, v_n)$ denote the vector of bids from all $n$ bidders.

\begin{proposition}\label{prop:single_dim_rev_scen1}
Assume that for each $i\in N$, $t_i=v_i e_i - \sum_{j \in N \sm i} \eta_{i
\gets j} e_j$ with $\eta_{i\gets j} =\alpha v_i$ for $j\in  N\sm i$. For bidder
$i\in N$, define their virtual value function $\phi_i(v)\eqdef v - \big(1-F_{v_i}(v)\big)/f_{v_i}(v)$ and assume it is non-decreasing.

Then, the allocation that maximizes revenue among BNIC and IR mechanisms is
\begin{align*}
	x_i(\bv)= \ind*{\phi_i(v_i)\geq \alpha\sum\nolimits_{j\in N \sm
	i}\phi_j(v_j)},\ \text{for $i\in N$ and $\bv \in \prod_{j \in N}[\ubar{v}_j, \bar{v}_j]$},\\
	x_j(v_i=\emptyset, \bv_{-i}) = \ind{j\neq i},\ \text{for $(i,j)\in N^2$ and  $\bv_{-i} \in \prod_{j \in N\sm i}[\ubar{v}_j, \bar{v}_j]$},
\end{align*}

An optimal payment implementing this allocation rule is given by
\begin{displaymath}
p_i(\bv) = \tau_{ii}(\bv_{-i})x_i(\bv)
+\alpha\sum_{j\in N\sm i} \tau_{ij}(\bv_{-i}) (1-x_j(\bv)),
\end{displaymath}
where $\tau_{ii}$ and $\tau_{ij}$ are the threshold types defined by
\begin{displaymath}
\tau_{ii}(\bv_{-i}) \eqdef \phi_i^{-1}\prn[\bigg]{\alpha\sum_{j\in N\sm i}\phi_j(v_j)}
\quad\text{and}\quad
\tau_{ij}(\bv_{-i}) \eqdef \phi_i^{-1}\prn[\bigg]{\phi_j(v_j)/\alpha
- \sum_{\mathclap{k\in N\sm \set{i,j}}}\phi_k(v_k)}.
\end{displaymath}
\end{proposition}

The proof of \cref{prop:single_dim_rev_scen1} is given in
\cref{appendix:prop:single_dim_rev_scen1}. Under
the assumption that $\eta_{i\gets j} = \alpha v_i$, the interim allocations
$\by^{(i)}$ and interim payment $q_i$ can be written as a function of $v_i$
only. Furthermore, bidder $i$'s interim utility becomes linear in the single term $a_i(v_i)\eqdef y^{(i)}_i(v_i)-\alpha\sum_{j\in N \sm i}y^{(i)}_j(v_i)$, which represents the interim aggregated effect of allocations on bidder $i$: 
\begin{displaymath}
	\cex{u_i\big(\bx(\bt),\bp(\bt);\bt\big)}{t_i} = v_i\cdot a_i(v_i) - q_i(v_i).
\end{displaymath}
Hence, the situation
becomes equivalent, from the perspective of bidder $i$, to the auction of
single item with linear utility. The characterization of BNIC therefore
simplifies (see \cref{lemma:ic-1d}) and becomes essentially equivalent to the
one of \cite{M81}, only requiring that $a_i$ be non-decreasing and allowing us
to write the expected revenue, up to an additive constant as
\begin{displaymath}
	\ex*{\sum_{i\in N}x_i(\bt)\left(\phi_i(v_i)-\alpha\sum\nolimits_{j\neq i}\phi_j(v_j)\right)}.
\end{displaymath}
From here, the optimal allocation is easily obtained and the derivation of the payment is analogous to the one in \cref{cor:restricted-dep-cor}.

\subsection{Revenue Maximization in Scenario 2: Known Outgoing Externalities}\label{sec:opt-rev-1}

Recall that in Scenario 2, the private type of each bidder $i \in N$ is $t_i = v_i e_i - \sum_{j \in N \sm i} \eta_{j \gets i} e_j$.
Using the BNIC characterization of \cref{prop:BNIC_char}, \cref{prop:rev} below shows that the problem of finding the revenue-optimal mechanism reduces to solving $n$ distinct optimizations over univariate functions.
Throughout this section, we denote by $F_{v_i}$ (resp.\ $f_{v_i}$) the cumulative (resp.\ probability) distribution function of the marginal distribution of $v_i$, for $i\in N$.
\begin{theorem}\label{prop:rev}
Suppose bidders have private types of the form $t_i = v_i e_i - \sum_{j \in N \sm i} \eta_{j \gets i} e_j $ for $i \in N$. Define the virtual valuation functions $ \phi_i(v_i) \eqdef
	v_i - \big(1-F_{v_i}(v_i)\big)/f_{v_i}(v_i) $ for $i \in N$. Suppose the functions $\widetilde\phi_i: v_i\mapsto\phi_i(v_i) -\sum_{j\in N\sm	i}\E[\eta_{j\gets i}\,|\, v_i]$ are non-decreasing, and define the thresholds $\tau_i \eqdef \widetilde\phi_i^{-1}(0)$.
 Then the mechanism with allocation functions
\begin{align*}
	x_i(\bt)&= \ind*{v_i\geq \tau_i},\ \text{for $i\in N$ and $\bt\in T$},\\
	x_j(t_i=\emptyset, \bt_{-i}) &= \ind{j\neq i},\ \text{for $(i,j)\in N^2$ and $\bt_{-i}\in T_{-i}$},
\end{align*}
 and payments 
\begin{displaymath}
 p_i(\bt) =x_i(\bt)\cdot\tau_i +\sum_{j\in N\sm i}
	\E\big[\eta_{i\gets j}\big(1-x_j(\bt)\big)\big],\ \text{for $i\in N$},
\end{displaymath}
is revenue-optimal among all BNIC and interim IR mechanisms.
\end{theorem}

The proof of \cref{prop:rev} can be found in \cref{appendix:rm2_proof}. Similar to the single parameter setting \citep{M81}, we obtain that under the stated regularity assumption, the optimal allocation rule takes a simple form: set a threshold value for each bidder above which all data is allocated and below which no data is allocated. In other words, the optimization problem reduces to finding $n$ parameters: the optimal threshold value of each bidder. 
Again, observe that $\widetilde\phi_i$ is similar in form to the threshold functions derived for the two social-welfare maximization cases \labelcref{eq:sw1_x,eq:sw2_expostalloc}.
In contrast to \cref{prop:NEWrm2_BNIC}, the virtual function is only applied to the value parameter $v_i$.

An interpretation of the payment rule is that each bidder $i$ pays their threshold value $\tau_i$ whenever they receive the good. Additionally, they pay an amount equal to the expected sum of externalities averted, which occurs when any bidder $j\in N \sm i$ is not allocated the good.

\begin{remark}
    Note that in the present case of known outgoing externalities, the optimal allocation rule only depends on the conditional expectations of the externality parameters. Intuitively, since each buyer $i \in [N]$ reports parameters $\set{\eta_{j \gets i}}_{j\neq i}$ which do not enter into their own utility but do affect the allocations to others, the BNIC condition (see Proposition \ref{prop:BNIC_char}) can only be enforced if these reports are ignored. This condition was also relevant for the impossibility result for ex-post efficiency in Proposition \ref{prop:sw2opt-impossibility}.
    
    In contrast, in the case of known \emph{incoming} externalities studied in Section \ref{sec:rm1}, all parameters reported by a buyer enter their own utility and the seller is thus faced with a full-blown multidimensional screening problem, which is significantly more difficult. 
\end{remark}

\begin{example}\label{ex:rev}
In the special case where $(\eta_{j\gets i})_{j \in N \sm i}$ is independent of $v_i$, then $\E[\eta_{j\gets i}\,|\, v_i] = \E[\eta_{j\gets i}]$ and the last term in the definition of $\widetilde\phi_i$ does not depend on $v_i$. In this case our assumption on $\widetilde\phi_i$ is equivalent to the standard regularity assumption of the marginal distribution $F_{v_i}$ of $v_i$ (see \cite{M81}). The payments also take the simpler form:
	$
		p_i(v_i) = \ind{v_i\geq\tau_i}\cdot\tau_i + \sum_{j\in N\sm i}
		\E[\eta_{i\gets j}]\mathbb{P}[v_j< \tau_j]
		\,.
	$
\end{example}

\subsubsection{Single-dimensional types}
In analogy to \cref{sec:scen1-single-dim-type}, we now consider the setting of perfectly correlated type parameters, so that for each bidder $i \in N$ with value $v_i$ for the good, the externality bidder $i$ causes on every other bidder $j\in N\sm\set{i}$ is $\eta_{j \gets i} = \alpha v_i$, where $\alpha$ is a common and publicly known constant. The impact of a bidder's allocation is proportional to that bidder's own gains from their allocation.

Under this assumption, each bidder $i$'s type, $t_i = v_i e_i - \alpha v_i e_j$, is effectively one-dimensional.
We reject any bids $\hat{t}_i$ for which $\hat{\eta}_{j \gets i} \neq \alpha v_i$, so that valid bids are essentially reports of $v_i$, and once again let $\bv = (v_1, \dots, v_n)$ denote the vector of bids from all $n$ bidders.
In this setting, have that $\cex*{\eta_{j \gets i}}{v_i} = \alpha v_i$ for $j\in N\sm\set{i}$, and a direct application of \cref{prop:rev} yields the following corollary describing the optimal mechanism in this case.

\begin{corollary}\label{cor:single-dim-type-scen2}
Under Scenario 2, suppose bidders have one-dimensional types $v_i$ with
$\eta_{j \gets i} = \alpha v_i$ for each $i \in N$ and $j\in N\sm\set{i}$, and
that types are independent across bidders. Define the virtual value functions
$\phi_i(v_i) \eqdef v_i - \big(1-F_{v_i}(v_i)\big)/f_{v_i}(v_i)$ and suppose the the functions $\widetilde\phi_i:v_i \mapsto \phi_i(v_i) - (n-1)\alpha v_i$ are non-decreasing. Then the allocation the optimal mechanism is given by
\begin{align*}
x_i(\bv) = \tilde x_i(v_i) &\eqdef \ind{\phi_i(v_i) \geq (n-1) \alpha v_i}, \ \text{for } i \in N \text{and } \bv \in \prod_{j \in N}[\ubar{v}_j, \bar{v}_j] \\
x_j(v_i=\emptyset, \bv_{-i}) &= \ind{j\neq i},\ \text{for $(i,j)\in N^2$ and $\bv_{-i} \in \prod_{j \in N\sm i}[\ubar{v}_j, \bar{v}_j]$}.
\end{align*}
Defining the thresholds $\tau_i\eqdef \widetilde\phi_i^{-1}(0)$, a corresponding optimal payment function is given by
\begin{align*}
    p_i(\bv) =  \tilde{x}_i(v_i)\tau_i + \alpha \sum_{j \in N \sm i} \ex*{v_j (1-\tilde{x}_j(v_j)}.
\end{align*}
\end{corollary}

Observe that for any distribution of bidder types, when there are at least two bidders and $\alpha$ is large enough, the functions $\tilde{\phi}_i$ will no longer be non-decreasing. 
For example, consider the case when values are distributed uniformly in $[0,1]$, so that $\tilde{\phi}_i(v) = (2 - (n-1)\alpha) v - 1$. 
The mechanism provided in \cref{cor:single-dim-type-scen2} requires that
$\alpha \leq 2/(n-1)$, and works when the tradeoff between the value $v_i$
provided and the externalities $\eta_{j \gets i} = \alpha v_i$ caused by
allocating to bidder $i$ is manageable. Once $\tilde{\phi}_i$ is no longer
non-decreasing, however, the mechanism is no longer BNIC and bidders may be incentivized to misrepresent their types $v_i$ if a large $v_i$ reveals a large negative impact on other firms.

\section{Discussion of Results}\label{sec:summary}

\paragraph{Summary of Results.} 
In this paper, we introduced a model for
data auctions with $n$ firms in the presence of externalities. Our main
modeling assumptions reduce the (a priori) combinatorial complex problem associated
with allocating data to one of designing auctions for a single, freely-replicable
good, with linear utilities parameterized by $n$-dimensional types. Depending
on the form of the private types of the firms, we described revenue-maximizing and welfare-maximizing mechanisms and found that in all cases, under appropriate assumptions,
the optimal allocations is to either allocate all the data to a firm if its
value for data sufficiently outweighs the externalities it causes on other
firms, or otherwise allocate none of the data.
The specific form of this comparison depends on the situation considered and is summarized in \cref{table:summary}.

\begin{table}[h]
\begin{center}
\renewcommand{\arraystretch}{1.2}
\begin{tabular}{@{}r@{\hspace{1em}}l@{\hspace{1em}}l@{}}
	\toprule
        ${x_i(\bt) = \1\set{\cdot}}$
             & {Scenario 1 ($\be_{i \gets}$)}
             & {Scenario 2 ($\be_{\gets i}$)} \\
	\midrule
	{Welfare max.}
            & ${v_i \geq \sum_{j \in N \sm i} \eta_{j \gets i}}$
			& ${v_i \geq \sum_{j \in N \sm i} \E[\eta_{j \gets i}|v_i]}$ \\[0.85em]
			{Revenue max.\textsuperscript{\dag}}
            &  $\phi_{v_i}(v_i) \geq \sum_{j \in N \sm i} \phi_{\eta_{j \gets i}}(\eta_{j \gets i})$ 
            & $\phi_{v_i}(v_i) \geq \sum_{j \in N \sm i} \E[\eta_{j \gets i}|v_i]$ \\
            [0.7em]
			\parbox[c]{9em}{\raggedleft Revenue max.\ with\\ single dim.\ types}
            &  $\phi_{v_i}(v_i) \geq \alpha \sum_{j \in N \sm i} \phi_{v_j}(v_j)$ 
            & $\phi_{v_i}(v_i) \geq (n-1)\alpha v_i$ \\
	\bottomrule
 \end{tabular}
\end{center}
 \caption{\label{table:summary}Summary of efficient and optimal allocation thresholds for the settings considered in this paper. Recall that $\phi_{v_i}(v) \eqdef v - (1-F_{v_i}(v))/f_{v_i}(v)$ and $\phi_{\eta_{i \gets j}}(\eta) \eqdef \eta + F_{\eta_{i \gets j}}(\eta)/f_{\eta_{i \gets j}}(\eta)$. \textsuperscript{\dag}The formula for revenue-maximization in Scenario~1 requires the restricted-dependency assumption (\cref{sec:restricted}).}
\end{table}
We now provide some interpretation for \cref{table:summary}.
In Scenario~1, we go from welfare maximization to revenue maximization with restricted-dependency mechanisms by simply replacing the types (value for data and externalities) with virtual types.
This exactly mimics what happens in the standard result \cite{M81}, where virtual values reduce the problem of maximizing revenue to that of maximizing welfare.
In Scenario~2, a similar reduction holds, but only the value for data needs to be transformed via the virtual function.
Intuitively, since the externalities reported by a firm do not appear in the firm's own utility but rather only affect other firms' utilities in this scenario, the optimal allocations must ignore these reports and instead rely on the prior distribution of externalities.
An analogous pattern holds for revenue maximization in the case of single-dimensional types, where the externality parameters are perfectly correlated with the value parameters. 

The form of the revenue-maximizing payment rules are summarized in \cref{table:summary_payments}. Because the payment rules are determined up to a constant by the thresholding allocation rules, the welfare- and revenue-maximizing payments $p_i$ for each bidder $i$ take similar forms.
As discussed below \cref{cor:restricted-dep-cor}, in conjunction with the threshold allocation rules, we see that the payment rules in Scenario~1 extend the intuition of second price auctions: bidders pay the minimum bid of $v_i$ needed in order to receive the good, plus the minimum report of externalities $\eta_{i \gets j}$ needed to prevent other bidders $j$ from receiving an allocation.
Both the welfare-maximizing VCG payments and the revenue-maximizing payment rules take this form. 
A similar interpretation holds in Scenario~2, where each bidder pays the minimum bid of $v_i$ needed to receive the good plus the expected externality that was averted from each bidder $j \in N \sm i$ who did not end up getting the good.
Since each bidder pays an amount equal to the potential externalities that were averted, even buyers that receive no data may have a positive payment.

\begin{table}[h]
\begin{center}
\renewcommand{\arraystretch}{1.2}
\begin{tabular}{@{}r@{\hspace{1em}}l@{\hspace{1em}}l@{}}
	\toprule
        ${p_i}$
             & {Scenario 1 ($\be_{i \gets}$)}
             & {Scenario 2 ($\be_{\gets i}$)} \\
	\midrule
	{General types\textsuperscript{\dag}}
            & ${x_i \cdot \tauii + \sum_{j \in N \sm i} (1-x_j)\cdot \tauij}$
			& ${x_i \cdot \tauii + \sum_{j \in N \sm i}  \E[(1-x_j)\cdot \eta_{i \gets j}]}$ \\
			{Single-dim. types}
            &  ${x_i \cdot \tauii + \alpha \sum_{j \in N \sm i} (1-x_j)\cdot \tauij}$ 
            & ${x_i \cdot \tau_i + \alpha \sum_{j \in N \sm i} \E[(1-x_j)\cdot v_j ]}$ \\
	\bottomrule
 \end{tabular}
\end{center}
 \caption{\label{table:summary_payments} Optimal payment rules for both settings of private types considered in this paper.\protect\footnotemark{} \textsuperscript{\dag}The formula for revenue-maximization in Scenario~1 requires the restricted-dependency assumption (\cref{sec:restricted})}
\end{table}
\noindent

\footnotetext{The thresholds $\tauii$ and $\tauij$ for bidder $i$ in Scenario 1 are such that the allocation functions in \cref{table:summary} can be written (with the stated monotonicity assumptions) $x_i(\bt) = \ind{v_i \geq \tauii}$ and $x_j(\bt) = \ind{\eta_{i \gets j} \leq \tauij}$ and depend on other bidders' bids. Likewise, the thresholds $\tau_i$ in Scenario 2 are such that $x_i(\bt) = \ind{v_i \geq \tau_i}$ for $i \in N$.}

\paragraph{Knowledge of the Type Distribution.} 
Note that the optimal mechanisms are parameterized by $n$ thresholds, determining the allocation and payment of each bidder.
In Scenario 2, these optimal thresholds depend only on the type distribution, the knowledge of which is thus required to run the optimal auctions.
When this distribution is unknown, a natural setting to consider is where the auctioneer learns the optimal thresholds over time by using a sequence of posted price mechanisms, adjusting the value of the thresholds at each time step based on the observed buyer's decisions.
We explore such a setting in \cref{sec:learning}. 

\paragraph{Mixed Information Structures.} 
While we focus on two information scenarios, one in which all incoming externalities are privately known, and the other in which all outgoing externalities are privately known, there may be situations in which a mixture of incoming and outgoing externalities are known. 
In general, extending our results to mixed types requires motivating a specific mixed type knowledge structure, a separate specification of what to do when type vectors overlap, and possibly adding assumptions for tractable BNIC characterizations and tractable revenue maximization.

First, if the type vectors of two bidders include the same type (e.g., both bidder 1 and bidder 2 know the externality caused by bidder 2 onto bidder 1) and assuming the type structure is known to the auctioneer, the auctioneer can take advantage of this perfect correlation of types and impose an arbitrarily large penalty if the two bidders report different values of that common type. Thus there will be a Bayes--Nash equilibrium in which the bidders report common types truthfully.
 
Next, if bidders' type vectors are disjoint in the sense that no bidder parameter is observed by more than one bidder, we can derive characterizations of Bayes Nash incentive compatibility similar to Proposition \ref{prop:JMS99BNICchar} and Proposition \ref{prop:BNIC_char}, with the additional standard assumption that bidder types vectors are independent across bidders. This assumption may not make much practical sense for arbitrary knowledge structures (e.g., if bidder 1 knows $v_1, \eta_{1\gets 2}$, and $\eta_{2 \gets 3}$ while bidder 2 knows $v_2, \eta_{1 \gets 3}$). Further, any parameter which is not reported will be considered in expectation conditional on the reported type vectors. 

Additionally, as long as one bidder has a private type that includes incoming externalities (and these parameters are not also present in another bidders' type to land in the first case), then the reduction for the impossibility result of Proposition \ref{prop:impossibility} still holds, showing that solving for the optimal mechanism is at least as hard as solving a multi-item, single-bidder auction. Additional assumptions for tractability may be imposed, as was done with the ``restricted dependency'' assumption in Section \ref{sec:restricted}, but again, whether these assumptions are sensible or natural depends on the specific mixed-type information structure.

\begin{figure}[!t]
\centering
\includegraphics[scale=0.5]{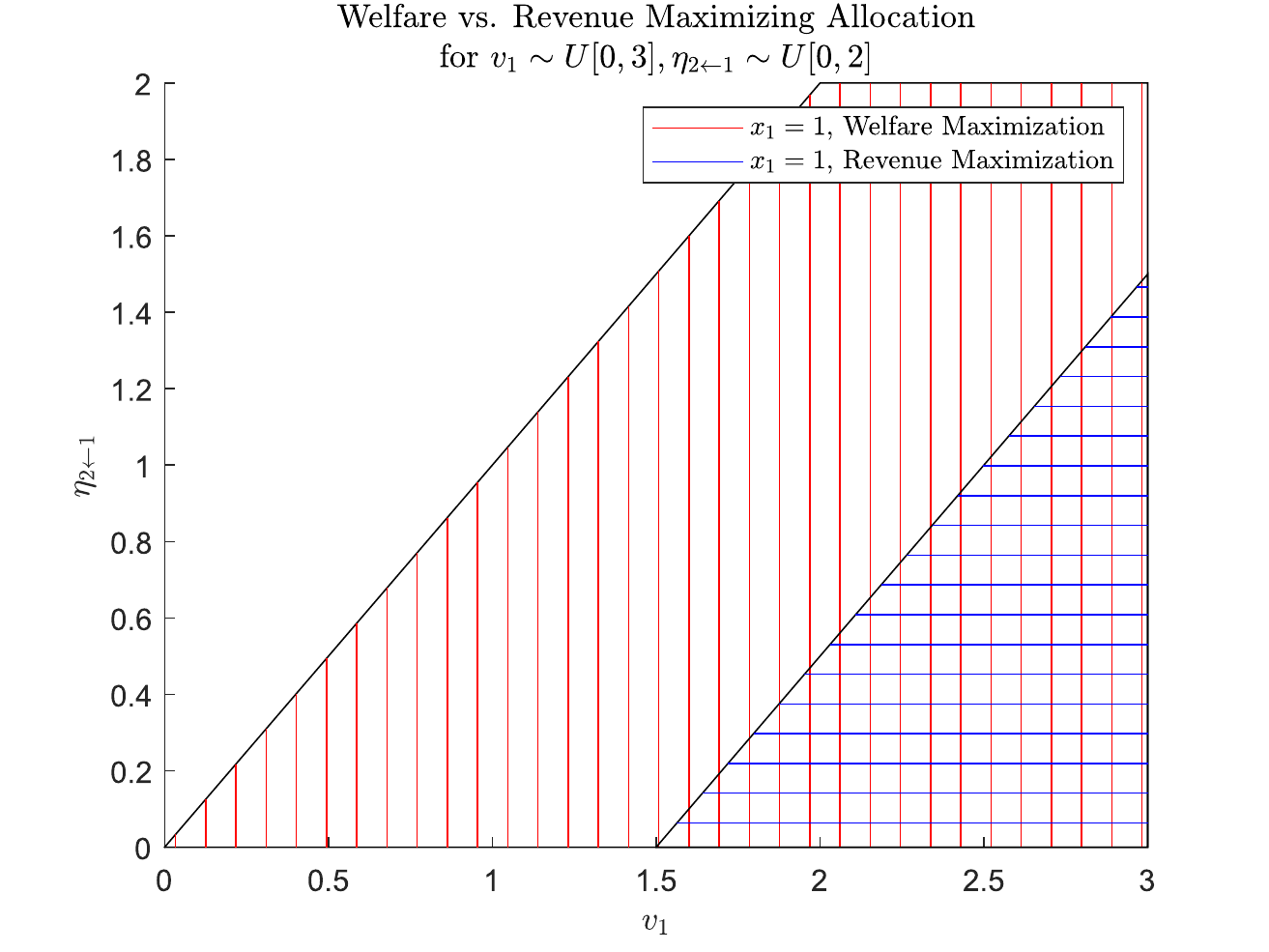}
\vspace{-1em}

\caption{\label{fig:type1_heatmap}Partition of type space by welfare versus revenue maximizing restricted dependency allocations in Scenario 1, assuming $v_1$ and $\eta_{2\gets 1}$ are uniformly distributed on their respective domains $[0, 3]$ and $[0,2]$. The shaded regions denote where bidder 1 is allocated the entire dataset ($x_1 = 1$) and the un-shaded regions correspond to the opposite case of $x_1 = 0$.}
\end{figure}

\paragraph{Welfare vs. Revenue Maximization.} To provide intuition on the differences between efficient and optimal mechanisms, consider
the special case of two bidders with uniformly distributed type parameters in
Scenario 1. The revenue-maximizing restricted dependency allocation function allocates to bidders less often
than does the welfare-maximizing allocation function and is in general not efficient.
This is illustrated in \cref{fig:type1_heatmap}, where the
welfare-maximizing and revenue-maximizing allocations are shown to partition
the type space of $\bt$ into the regions based on bidder 1's allocation. For
details, see \cref{appendix:heatmap_type1}.
Keep in mind that these results are obtained under different assumptions. The social
welfare case in Scenario 1 is an instantiation of the VCG mechanism and
requires no assumption beyond our externality model. In Scenario 2, since firms
do not know the externality other firms cause on them, they have to reason in
expectation about their utility and hence this scenario requires a common known
prior on the type distribution.

\paragraph{Impact of Externalities.}
To further elucidate the effect of the presence and magnitude of negative externalities on the optimal revenue, we consider the simple setting where each bidder $i \in N$ has value $v_i$ distributed uniformly in $[0, \bar{v}_i]$, while externality parameters $\eta_{i \gets j}$ are deterministic, for $j \in N \sm i$. 
Since the externalities are now common knowledge, Scenario~1 and Scenario~2 are identical. In particular, the optimal mechanism described in \cref{prop:rev} and takes a simple form that we now describe.

For uniform type distributions, the virtual valuation functions $\phi_i(v_i)$ are non-decreasing in $v_i$.
A simple computation shows the expected optimal revenue can thus be expressed as $\sum_{i \in N} R_i$, where $R_i$ denotes the share of each bidder's payment induced by the presence of bidder $i$, and is given by
\begin{align}\label{eq:rev_vs_ext}
R_i = \begin{cases}
 \sum_{j \in N \sm i} \eta_{j \gets i} &\text{if }  \sum_{j \in N \sm i} \eta_{j \gets i} > \bar{v}_i,
\\[8pt] \frac{(\bar{v}_i -  \sum_{j \in N \sm i} \eta_{j \gets i})^2}{4\bar{v}_i} +  \sum_{j \in N \sm i} \eta_{j \gets i} &\text{if }  \sum_{j \in N \sm i} \eta_{j \gets i} \leq \bar{v}_i.
\end{cases}
\end{align}
Some of comments about this expression are in order:
\begin{enumerate}[leftmargin=*]
	\item When there are no externalities ($\eta_{j\gets i} =0$ for $j\in N\sm
		i$), we recover the revenue of the optimal posted price mechanism $\bar
		v_i/4$. Otherwise, we see that $R_i$---and thus the overall
		revenue---is increasing in the externality parameters $\eta_{j\gets i}$
		for all $i,j\in N$ with $i\neq j$.
	\item Without externalities, the optimal posted price mechanism allocates
		with probability $1/2$. In contrast, in the presence of externalities,
		there are two regimes: if $\sum_{j\neq i} \eta_{j\gets i}\leq \bar
		v_i$, the optimal mechanism allocates to $i$ with probability $1/2
		- \sum_{j\neq i} \eta_{j\gets i}/(2\bar v_i)$, otherwise it never
		allocates.
	\item In both regimes---in particular, even when bidder $i$ is not allocated---the optimal mechanism is able to collect at least $\sum_{j\in N\sm i}\eta_{j\gets i}$, corresponding to the ``threat'' of allocating to bidder $i$ in the outside option where bidders do not participate.
\end{enumerate}

\begin{figure}[!tb]
\centering
\includegraphics[scale=0.45]{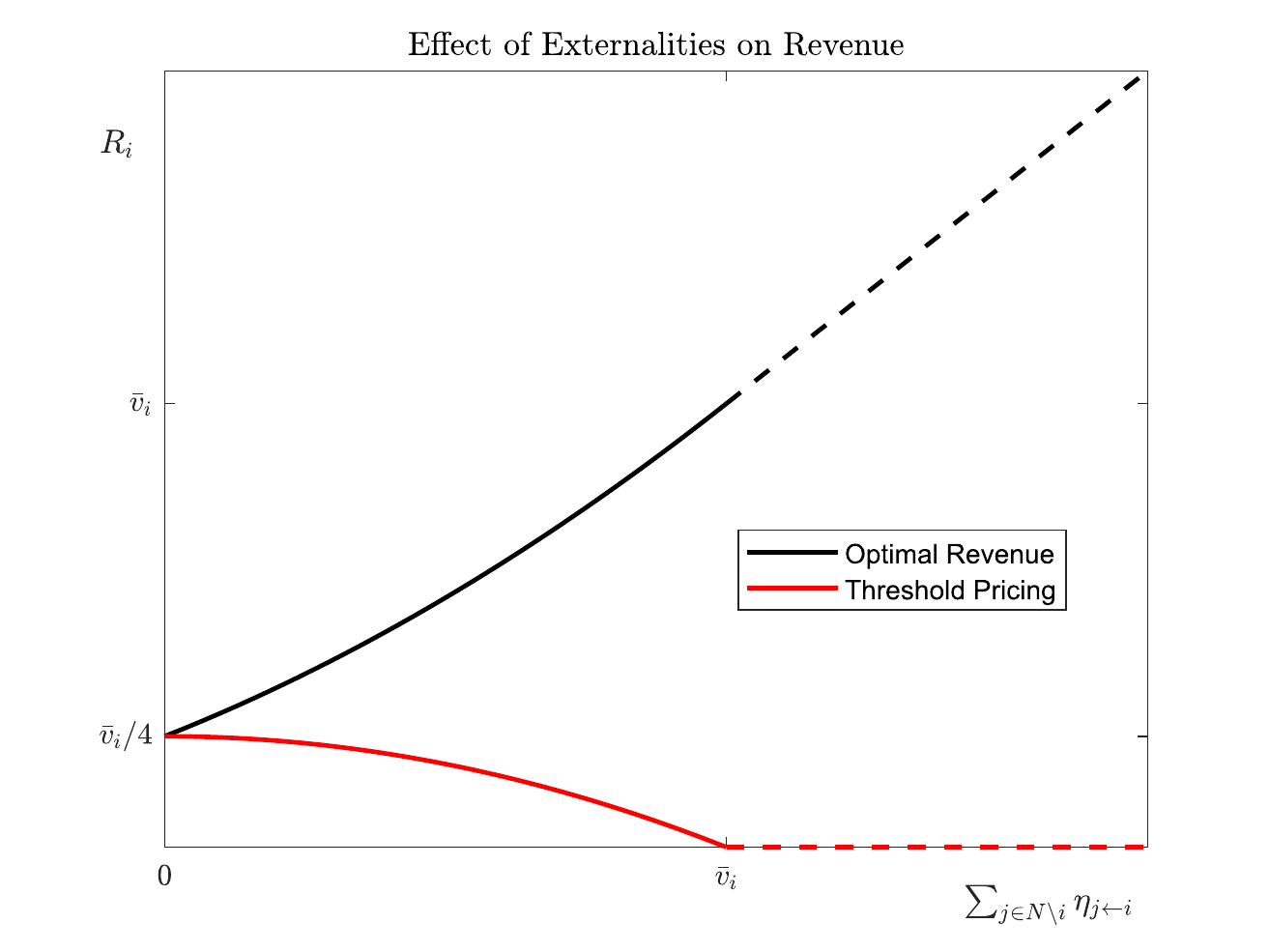}
\vspace{-1.2em}
\caption{\looseness=-1 Contribution $R_i$ to total revenue due to the presence of bidder $i$
	as a function of the externality caused by bidder $i$: $\sum_{j\in N\sm
	i} \eta_{j \gets i}$. In black, the value of $R_i$ in the optimal mechanism, given by
	Equation~\eqref{eq:rev_vs_ext}. In red, a suboptimal mechanism
	that only charges the optimal threshold $\tau_i$ when allocating to bidder
	$i$, without extracting additional revenue from bidders $j\neq i$ corresponding to the
threat of their outside option.}
\label{fig:rev_v_ext}
\end{figure}

In summary, the presence of externalities implies that the auctioneer can still extract payments from bidders, even when they do not receive an allocation, while still maintaining IR. 
It turns out that an auctioneer benefits from greater externalities among bidders: even though increased externalities may lead to fewer allocations and therefore less payments directly driven by these allocations (based on $v_i$), the entry fee that the auctioneer charges %
can make up for and actually exceed this loss in profit. However, if the auctioneer were to only charge the optimal threshold $\tau_i = (\bar v_i+
\sum_{j\neq i} \eta_{j\gets i})/2$ upon allocating, \emph{without} leveraging the outside option, the revenue would in fact be decreasing in the externality
parameters. These phenomena are illustrated in \cref{fig:rev_v_ext}.

\subsection*{Acknowledgments}
The authors are grateful to Dirk Bergemann, Alessandro Bonatti, Tan Gan, Andreas Haupt, Ali Jadbabaie  and Haifeng Xu for fruitful discussions and comments about this work.

\bibliographystyle{plainnat}
\bibliography{main}

\appendix
\section{Prior-independent mechanism}\label{sec:learning}

A natural concern with the optimal mechanisms described in
Section~\ref{sec:RM} is that they rely on the knowledge of the type
distribution. For example, even in the \emph{regular} case of
Theorem~\ref{prop:rev}, the threshold value $\tau_i$ of bidder $i$ is
described as the zero of the virtual value function which itself depends
heavily on the distribution of the type of bidder $i$. In this section, we
explore a situation where the type distribution is initially unknown by the
mechanism designer and learned over time.

This problem falls within the realm of online stochastic optimization, and more specifically stochastic derivative-free optimization: indeed, the binary decisions of the bidders at each time step let us compute an unbiased estimator of the the expected revenue, which is the only prerequisite of derivative-free optimization procedures. Standard procedures to do so, such as the one described in \cite{FKM05}, can then be applied to algorithmically update the thresholds at each time step and provide an upper bound on the rate of convergence to the optimal revenue.
\subsection{Problem description}
\label{sec:learning-problem}

We consider a sequential learning setting in which at each time step, the
seller attempts to sell a digital good in an auction with $n$ buyers whose
types are drawn from an unknown distribution. The task of the seller is to
design an auction mechanism to both satisfy the BNIC and IR constraints at each
time step, and also learn the unknown distribution over time and converge to
the optimal auction. Specifically, at time step $k\geq 1$:
\begin{enumerate}
	\item the data seller announces the BNIC and IR mechanism $\mu^k
		= \big(\bx^k(\cdot), \bp^k(\cdot))$
		which will be used for this time step;
\item a vector of types $\bt^k\in T$ is drawn from the distribution $f$
	independently of previous time steps;
\item the good is allocated according to $\bx(\bt)$ and the seller
	collects the payments $\bp(\bt)$.
\end{enumerate}

We emphasize that the underlying distribution $f$ from which the types are
drawn is unknown to the seller and so $\mu^k$ cannot depend \emph{directly} on
it. However, the seller observes the types reported in previous time steps and
can use these observations to inform the design of $\bx^k(\cdot)$ and
$\bp^k(\cdot)$. A natural benchmark to evaluate the performance of the chosen
sequence of mechanisms is to compare its cumulative revenue to the one which
would have been obtained using the optimal auction.  Formally, let
$\bx^\star(\cdot)$ and $\bp^\star(\cdot)$ denote the revenue-optimal BNIC and
IR auction---which depends on the distribution $f$---and define the regret up
to time step $K$ by
\begin{displaymath}
	R(K) = \sum_{k=1}^K \big[ \rev(\bx^\star, \bp^\star)
		-\rev(\bx^k,\bp^k) \big]
		\,,\quad K\geq 1
		\,.
\end{displaymath}

The goal of the data seller is thus to design a sequence of mechanisms which
incurs as small a regret as possible.

\begin{remark}
An important assumption in this sequential learning setting is that the buyers'
types are redrawn at each time step from the same distribution and that buyers
only strategize within each time step (hence the BNIC and IR constraints) but
not \emph{across} times steps. As such, we ignore intertemporal incentive
issues, which could for example lead buyers to misreport their types in order
to bias the learning of the distribution that the seller performs over time.
We note that an alternative learning model would consider a single-shot auction
in which buyers' types are drawn from identical distributions. In this case, one
can still hope to learn the type distribution in a single time step as the
number of buyers $n$ goes to infinity. The problem of buyers misreporting their
types can be addressed by learning the unknown distribution from a subsample
$S\subset N$ of the buyers and then using the learned mechanism only on the
remaining $N\sm S$ buyers. This approach was for example adopted in
\cite{GHKSW06}.
\end{remark}

\subsection{A stochastic optimization solution}

We now provide a solution to the sequential learning problem of
Section~\ref{sec:learning-problem} which leverages the specific structure of
the optimal mechanism described in Section~\ref{sec:opt-rev-1}. We will make
the following two simplifying assumptions:
\begin{itemize}
	\item for each buyer $i$, the distribution of her value $v_i$ is
		independent from the distribution of externalities $\be_{\gets i}$ she causes on others. Furthermore, for all $i$ and
		$j\in N\sm i$, the expectation $\E[\eta_{i\gets j}]$ is known to the
		seller;
	\item for each buyer $i$, the function $v_i\mapsto v_i f_{v_i}(v_i) - (1-
		F_{v_i}(v_i))
		- f_{v_i}(v_i)\sum_{j\in N\sm i} \E[\eta_{j\gets i}]$ is non-decreasing.
\end{itemize}
Under these assumptions, Theorem~\ref{prop:rev} combined with
Example~\ref{ex:rev} imply that the revenue-optimal BNIC and IR mechanism is
given by
\begin{displaymath}
	x_i^\star(\bt) = \ind{v_i\geq \tau_i^\star}
	\quad\mathrm{and}\quad
	p_i^\star(\bt) = \ind{v_i\geq\tau_i^\star}\cdot\tau_i^\star + \sum_{j\in N\sm i}
	\E[\eta_{i\gets j}]\cdot F_{v_j}(\tau_j^\star)
		\,,
\end{displaymath}
where $\tau_i^\star \eqdef \widetilde\Phi_i^{-1}(0)$ for $\widetilde\Phi_i$
described in Theorem~\ref{prop:rev}. A simple computation further implies that
\begin{equation}
	\label{eq:objective}
	H(\btau^\star)
	\eqdef \rev(\bx^\star,\bp^\star)
	= \sum_{i\in N}\left[\tau_i^\star
	-F_{v_i}(\tau_i^\star)\bigg(\tau_i^\star - \sum_{j\in N\sm i}\E[\eta_{j\gets i}]\bigg)\right]
	\,.
\end{equation}

This formula has a number of important consequences. First, observe that the
mechanism obtained by replacing $\btau^\star$ by any other vector of thresholds
$\btau$ is also BNIC and IR. Since the optimal mechanism is given by the choice
of $\btau^\star$, optimizing the function $H$ over the choice of $\btau\in
\prod_{i\in N} [\ubar v_i, \bar v_i]$ is sufficient to obtain the optimal
mechanism. In other words, the problem of finding the optimal mechanism reduces
to the optimization of the function $H$. Furthermore, note that this
maximization problem is \emph{separable}: denoting by $H_i$ the summand in
\eqref{eq:objective} we have that $H(\btau) = \sum_{i\in N} H_i(\tau_i)$ and
hence, the problem reduces to separately maximizing each function $H_i$ over
the choice of a single-dimension parameter $\tau_i$. Finally, observe that the
derivative of $H_i$ is $h_i(\tau_i)\eqdef 1 - \tau_ig_i(\tau_i)
- G_i(\tau_i)-g_i(\tau_i)\sum_{j\in N\sm i}\E[\eta_{j\gets i}]$ which is
non-increasing by assumption. Hence, the function $H_i$ is concave.

\paragraph{Learning as stochastic optimization.}

The previous observations suggest a natural strategy to learn the optimal
auction in the observational model described in
Section~\ref{sec:learning-problem}: \emph{using a sequence of posted price
mechanisms and iteratively updating the thresholds based on the buyers'
decisions}.

Note that when using a posted price mechanism with threshold $\tau_i^k$ at time
step $k$, the only thing we observe from buyer $k$ is her decision to accept or
reject the offer, which is the binary random variable
$X_i^k\eqdef\ind{v_i^k\geq \tau_i^k}$. Hence, the suggested strategy does not
even require the buyers to report their entire type but simply relies on the
binary observations $(X_i^k)_{i\in N}$ from each time step. From these
observations, one can form for each $i\in N$, the quantity
\begin{displaymath}
	H_i(\tau_i^k, X_i^k) \eqdef \tau_i^k - (1-X_i^k)\bigg(\tau_i^k - \sum_{j\in N\sm i}
	\E[\eta_{j\gets i}]\bigg)
\end{displaymath}
which is an unbiased estimate of $H_i(\tau_i^k)$, that is, $H_i(\tau_i^k)
= \E\big[H_i(\tau_i^k, X_i^k)]$.

\paragraph{}
This observation conveniently reduces the problem of learning the optimal mechanism to the one of maximizing for each $i\in N$ the concave function $H_i(\cdot)$ assuming access
to unbiased estimates of $H_i(\tau_i^k)$ for a sequence $\tau_i^k$ of our
choice. This problem is exactly what is known in the online learning literature
as a \emph{stochastic derivative-free optimization problem} where the quantity
$H_i(\tau_i^k, X_i^k)$ plays the role of a 
\emph{zeroth order stochastic oracle}. (The stochastic
optimization literature distinguishes between first-order stochastic oracles giving unbiased estimates of the gradient of the objective function and zeroth-order oracles given unbiased estimates of the objective function
itself.)

\paragraph{}
Algorithms for stochastic derivative-free optimizations usually take the form
of iterative algorithms in which at each time step, the current estimate of the
optimal solution is updated based on the previous estimate and the current
observation. In our notations, we write:
\begin{displaymath}
	\tau_i^{k+1} = \mathrm{Update}(\tau_i^k, X_i^k)
\end{displaymath}
where $\mathrm{Update}(\cdot)$ is the update rule given by a stochastic optimization
algorithm. We instantiate the stochastic optimization framework in our setting
and describe in Algorithm~\ref{alg} a meta algorithm parametrized by a generic
$\mathrm{Update}$ function.

\begin{algorithm}[t]
	\KwIn{vector of initial thresholds $\btau^1$, $\mathrm{Update}$ given
	a stochastic optimization algorithm}
	\For{$k\geq 1$}{
		post vector of thresholds $\btau^k$\;
		collect buyers' decisions $X_i^k = \ind{v_i^k\geq \tau^k_i}$\;
		\For{$i\in N$}{
			\eIf{$X_i^k==1$}{
				allocate good\;
				collect payment $\tau_i^k + \sum_{j\in N\sm i}\E[\eta_{i\gets j}](1-X_j^k)$\;
				}{
				do not allocate good\;
				collect payment $\sum_{j\in N\sm i}\E[\eta_{i\gets j}](1-X_j^k)$\;
			}
			$\tau_i^{k+1}\gets\mathrm{Update}(\tau_i^k, X_i^k)$\;
		}
	}
	\caption{A meta online learning algorithm from single-dimensional
		stochastic optimization}
		\label{alg}
\end{algorithm}

\paragraph{}
All that remains to do to obtain a concrete bound on the regret of our learning
procedure is to choose an $\mathrm{Update}(\cdot)$ function from the stochastic
optimization literature. Here, we will not aim for optimal bounds but rather
give a simple example of how to instantiate Algorithm~\ref{alg}. A common idea
is to estimate the derivative of the objective $H_i$ at $\tau_i$ by $H_i(\tau_i
+ \delta u) u/\delta$ where $u$ is uniformly random over $[-1,1]$. Using this estimate
of the derivative in projected gradient-ascent suggests the following simple
update rule:
\begin{equation}
	\label{eq:update}
	\tau_i^{k+1} = \tau_i^{k} + \gamma\cdot H_i(\tau_i^k + \delta u, X_i^k)u/\delta
	\,,
\end{equation}
where $u$ is uniformly random over $[-1,1]$ and $\gamma$ is the \emph{step
size} parameter. By applying a standard result (see e.g. \cite{FKM05}) we
obtain the following proposition.

\begin{proposition}
	\label{thm:regret}
	For any $K\geq 1$, there exists a choice of $\delta$ and $\gamma$ such that
	using the update rule \eqref{eq:update} in Algorithm~\ref{alg} yields
	a regret $R(K) = O(K^{3/4})$.
\end{proposition}

Furthermore, since the revenue is a concave function of the chosen threshold,
we recover the standard fact that a bound on the regret implies a bound on the
objective function.

\begin{corollary}
	Consider the sequence of threshold vectors $(\btau_k)_{k\geq 1}$ computed
	by Algorithm~\ref{alg} with update rule \eqref{eq:update} and define
	$\overline\btau_K \eqdef \frac{1}{K}\sum_{k=1}^K\btau_k$ for $K\geq 1$, and
	denote by $\rev(\overline\btau_K)$ obtained by using the posted price
	mechanism with thresholds $\overline\btau_K$. Then for any $K\geq 1$, there
	exists a choice of $\delta$ and $\gamma$ such that:
	\begin{displaymath}
		\rev(\btau^\star) - \rev(\overline\btau_K)
		= O\left(\frac{n}{K^{1/4}}\right)
		\,.
	\end{displaymath}
\end{corollary}
\begin{proof}{}
{\color{white}.} This follows immediately from Proposition~\ref{thm:regret} by application of
Jensen's inequality.
\end{proof}

\section{Linear Externality Models} 
\subsection{Proof of Fact~\ref{fact:ex_quality}}\label{appendix:ex_quality}
\begin{proof}{}
	By assumption $(p_1-p_2)/(\beta_1q_1-\beta_2q_2)\geq\ubar{\alpha}>0$, which
	implies in particular that $p_1-p_2$ and $\beta_1q_1-\beta_2q_2$ have the
	same sign. Without loss of generality, assume that $p_1>p_2$ and
	$\beta_1q_1>\beta_2q_2$, otherwise the following proof applies after
	swapping the role of 1 and 2.

Given the distribution of consumer types,
\begin{align*}
    u_1(q_1, q_2) &= (p_1 - c_1) \mathbb{P}\set{\alpha: v_1(\alpha) > v_2 (\alpha)} \\ 
				  &= (p_1 - c_1) \ubar{\alpha}(\beta_1 q_1 - \beta_2 q_2)/(p_1 - p_2). 
\end{align*} 
Then taking $\nu_1(x_1, x_2)= u_1(q_1+x_1, q_2 + x_2) - u_1(q_1, q_2)$ and
noting the linearity of $u_1$ in its arguments, we obtain the result for firm
1. The result for firm 2 follows in a similar manner after observing that
   $u_2(q_1, q_2) = (p_2 - c_2) (1 - \mathbb{P}\set{\alpha: v_1(\alpha) > v_2
   (\alpha)})$.
\end{proof}

\subsection{Proof of Fact~\ref{fact:ex_cournot}}\label{appendix:ex_cournot}
Let $\xi = (\alpha_1, \alpha_2, c_1, c_2, M)$ be the original set of market-relevant parameters before data is purchased and production efficiencies are improved, and let $\xi'=(\alpha_1',\alpha_2',c_1,c_2,M)$ be the market parameters after an allocation of data, such that $\alpha_i' = \alpha_i + x_i$. 
For $i\in\set{1,2}$, firm $i$'s profit is given by
\begin{displaymath}
	\pi_i(q_1, q_2; \xi)
	= (M- \alpha_1\cdot q_1 - \alpha_2\cdot q_2)\cdot \alpha_i \cdot q_i - c_i \cdot q_i.
\end{displaymath}
For market parameters $\xi$, a standard computation gives the equilibrium production quantity and profits of firm 1:
\begin{align*}
q_1^*(\xi) = \frac{\alpha _{1}c_{2}-2\alpha _{2}c_{1}+M\alpha _{1}\alpha
_{2}}{3{\alpha _{1}}^2\alpha _{2}}
\quad\text{and}\quad
\pi_1^*(\xi)
= (\alpha_1 q_1^*)^2\,,
\end{align*}
with analogous expressions for those of firm 2 obtained by swapping indices 1 and 2.

We obtain by Taylor expansion the change in equilibrium profits 
\begin{equation}
	\label{eq:taylor}
	\Delta\pi_i(\xi',\xi)
	\eqdef \pi_i^*(\xi') - \pi_i^*(\xi)
	= \sum_{j = 1}^2 \ddp{\pi_i^*}{\alpha_j}(\xi)
 \cdot (\alpha_j' - \alpha_j) + O\big(\|\xi'-\xi\|^2\big)\,.
\end{equation}
Furthermore, a simple computation gives
$\ddp{\pi_1^*}{\alpha_1}(\xi) = \frac{4 c_1}{3 \alpha_1} q_1^*(\xi)$ and
$\ddp{\pi_1^*}{\alpha_2}(\xi) = -\frac{2\alpha _{1}c_{2}}{3{\alpha _{2}}^2}
q_1^*(\xi)\,.$ Assuming second order terms in \eqref{eq:taylor} are sufficiently small to be ignored, we obtain 
\begin{displaymath}
	\Delta\pi_1(x_1, x_2) = 
\frac{4 c_1}{3 \alpha_1} q_1^*(\xi)\cdot x_1
-\frac{2\alpha _{1}c_{2}}{3{\alpha _{2}}^2} q_1^*(\xi)\cdot x_2
=v_1\cdot x_1 - \eta_{1\gets 2}\cdot x_2\,,
\end{displaymath}
\begin{equation}
	\label{eq:firm_params_proof}
	\text{where}\quad v_1\eqdef 4c_1q_1^*(\xi)/(3\alpha_1)
\quad\text{and}\quad
\eta_{1\gets 2}\eqdef 2\alpha_1 c_2q_1^*(\xi)/(3\alpha_2^2)\,,
\end{equation}
and similarly for firm 2. This has the exact same form as
\eqref{eq:firm_valuation_model}. In other words, our externality model captures
the first order approximation of the change of profit that results from firms
acquiring data. This immediately extends to the general case of $n$ firms
engaging in Cournot competition.

\section{Welfare versus Revenue Maximization in Scenario 1} \label{appendix:heatmap_type1}
As stated in \cref{sec:rm1}, given a distribution function $F_{\eta_{i \gets j}}$ and corresponding density function $f_{\eta_{i \gets j}}$ for the random variable $\eta_{i \gets j}$ supported on $[\ubar{\eta}_{i \gets j}, \bar{\eta}_{i \gets j}]$, for $i\neq j \in N$, we define the distribution of $t_{i,j}$ on $[-\bar{\eta}_{i \gets j}, -\ubar{\eta}_{i \gets j}]$ by the distribution and density functions 
\begin{align*}
& F_{i,j}(t_{i,j}) = 1 - F_{\eta_{i \gets j}}(-t_{i,j}) = 1 - F_{\eta_{i \gets j}}(\eta_{i \gets j})
\\ & f_{i,j}(t_{i,j}) = f_{\eta_{i \gets j}}(-t_{i,j}) = f_{\eta_{i \gets j}}(\eta_{i \gets j}).
\end{align*} 
Further, define for $i, j \in N$ the virtual value function $\Phi_{i,j}(t_{i,j}) \eqdef t_{i,j} - \big(1-F_{i,j}(t_{i,j})\big)/f_{i,j}(t_{i,j})$. Then for each $i \in N$ and $j \in N \sm i$, we can express the virtual functions as 
\begin{align*}
& \Phi_{i,i}(t_{i,i}) = v_i - \big(1-F_{v_i}(v_i)\big)/f_{v_i}(v_i)
\\ & \Phi_{i,j}(t_{i,j}) = - \eta_{i \gets j} - F_{\eta_{i \gets j}}(\eta_{i \gets j})/f_{\eta_{i \gets j}}(\eta_{i \gets j})
\end{align*}

Suppose all $v_i, \eta_{i \gets j}$ for $i\in N, j \in N \sm i$ are uniformly distributed on their respective domains. The virtual value functions take the forms
\begin{align*}
& \Phi_{i,i}(t_{i,i}) = 2v_i - \bar{v_i}
\\ & \Phi_{i,j}(t_{i,j}) = -2\eta_{i \gets j} + \ubar{\eta}_{i \gets j}
\end{align*}

The optimal allocation rule \eqref{eq:rm1_alloc} then becomes
\begin{align*}
x_k(\bt) & = \ind[\bigg]{\sum_{i \in N} \phi_{i,k}(t_{i,k})\geq 0}
	  \\ & = \ind[\bigg]{(2v_k - \bar{v}_k) + \sum_{i \in N \sm k} (-2\eta_{i \gets k} +
\ubar{\eta}_{i \gets k}) \geq 0}
	  \\ & = \ind[\bigg]{v_k - \sum_{i \in N \sm k} \eta_{i \gets k} \geq \frac{\bar{v}_k - \sum_{i \in N \sm k} \ubar{\eta}_{i \gets k}}{2}}
\end{align*}
In the case of $n = 2$ bidders, bidder~$1$'s allocation is
\begin{align*}
x_1(\bt) = \ind[\Big]{v_1 - \eta_{2 \gets 1} \geq \frac{\bar{v}_1 - \ubar{\eta}_{2 \gets 1}}{2}}
\end{align*}

Meanwhile, the welfare-maximizing allocation rule for bidder~1 is 
\begin{align*}
x_1(\bt) = \ind[\big]{v_1 - \eta_{2 \gets 1} \geq 0}.
\end{align*}
Thus, the revenue-maximizing allocation allocates to bidders less often than does the welfare-maximizing allocation. The optimal mechanism is therefore not efficient in general. This is illustrated in \cref{fig:type1_heatmap} in \cref{sec:summary}, where the type space for $\bt$ is partitioned in terms of the welfare-maximizing and revenue-maximizing allocation to bidder~1.

\section{Characterizations of IC and IR Mechanisms}\label{sec:IC_IR_char}
An important step towards elucidating the solution structure of the
welfare and revenue maximizing mechanisms is to obtain
a characterization of the IC and IR constraints. Since our valuation model has
the same form as the one in \cite{JMS96,JMS99}, we rely on the
characterizations found in these papers, and state them below for completeness.

\subsection{Characterizations in Scenario 1: Incoming Externalities}\label{sec:type1_char}
Recall that each bidder $i$'s type is the vector of the form $t_i = v_i e_i - \sum_{j \in N \sm i} \eta_{i \gets j} e_j \in \R^n$ where $e_k$ denotes the
$k$th vector of the standard basis.

For each bidder $i \in N$, with true type $t_i$ and bid $\hat{t}_i$, the overall interim expected allocation is defined as $\cex{\bx(\hat{t}_i, \bt_{-i})}{t_i} = \E_{\bt_{-i}}[x(\hat{t}_i, \bt_{-i})]$ and the interim expected payment as $\cex{p_i(\hat{t}_i,\bt_{-i})}{t_i} = \E_{\bt_{-i}}[p_i(\hat{t}_i, \bt_{-i})]$.
Under the given assumption of independent bidder types, the interim expected allocation and payment functions do not depend on bidder $i$'s true type $t_i$, so we write $\by^{(i)}(\hat{t}_i)\eqdef \E_{\bt_{-i}}[x(\hat{t}_i, \bt_{-i})]$ to indicate the interim expected allocation function and likewise, $q_i(\hat{t}_i)\eqdef \E_{\bt_{-i}}[p_i(\hat{t}_i, \bt_{-i})]$ for the interim expected payment.

Note that $\by^{(i)}$
is a vector field mapping $T_i$ to $[0,1]^n$.  Finally, for every bidder $i \in
N$, we define the critical type $\circt_i = 
\ubar v_i e_i - \sum_{j \in N \sm i} \ubar \eta_{i \gets j} e_j$, which will feature in the following IC and IR characterizations. Note $\circt_i$ consists of the type parameter values that are of the smallest magnitude. 

\begin{proposition}[{\cite[Proposition 1]{JMS99}}] \label{prop:JMS99BNICchar}
Suppose bidders' private types are of the form $t_i = v_i e_i - \sum_{j \in N \sm i} \eta_{i \gets j} e_j$
for each bidder $ i \in N$. Then the mechanism $(\bx,\bp)$ is BNIC
if and only if for each bidder $i\in N$:
\begin{enumerate}[label=(\roman*)]
	\item $\by^{(i)}$ is a conservative vector field\footnotemark 
	\item $\by^{(i)}$ is a monotone vector field, that is
	$\ip[\big]{s_i - t_i}{\by^{(i)}(s_i)-\by^{(i)}(t_i)} \geq
	0$ for all $s_i, t_i\in T_i$.
\item for each type $t_i\in T_i$, the interim payment is given by
\begin{align} \label{eq:bnic1_payment_app}
q_i(t_i) = \ip[\big]{\by^{(i)}(t_i)}{t_i}
- \int_{\circt_i}^{t_i} \by^{(i)}(s_i)\cdot ds_i - C_i
\,,
\end{align}
where $C_i$ is an arbitrary integration constant whose value sets
$V_i(\circt_i; \circt_i)$, the interim utility of bidder $i$ when its type is
$\circt_i = \ubar v_i e_i - \sum_{j \in N \sm i} \ubar \eta_{i \gets j} e_j$.
\end{enumerate}
\end{proposition} 

\footnotetext{A vector field $\by:\R^n \to \R^n$ is conservative if it can be expressed as the gradient of some scalar potential function $\phi:\R^n \to \R$. Equivalently, line integrals of $\by$ are independent of path taken between the endpoints.}

We provide the following characterization of interim IR for BNIC
mechanisms that maximize revenue. %

\begin{proposition}  \label{prop:IRchar_type1}%
Suppose private types are of the form $t_i = v_i e_i - \sum_{j \in N \sm i} \eta_{i \gets j} e_j$ for each bidder $i \in N$. Then a revenue-maximizing BNIC mechanism satisfies the
interim IR constraint $V_i(t_i;t_i) \geq V_i(\emptyset; t_i)$ if and only if
this condition is satisfied for the critical type $\circt_i
= \ubar v_i e_i - \sum_{j \in N \sm i} \ubar \eta_{i \gets j} e_j$. Further, the optimal mechanism sets $V_i(\circt_i; \circt_i) = -\sum_{j \in N \sm i} \ubar{\eta}_{i \gets j}$.
\end{proposition}

\begin{proof}{}
We first show that the optimal outside option when bidder $i$ does not
participate allocates the digital good to all remaining participants $N \sm i$.
We then show that it suffices to check that interim IR is satisfied for the
type $\circt_i$, and finally find the optimal value of the integration constant
$V_i(\circt_i; \circt_i)$.

\paragraph{Optimal outside option.}
The interim IR constraint is essentially a constraint on the values that the
constant $C_i = V_i(\circt_i;\circt_i)$ can take. That is, after plugging in
the form of the payment rule \eqref{eq:bnic1_payment_app}, interim IR can be
expressed as: 
\begin{displaymath}
\forall i \in N, \forall t_i \in T_i\::\:
V_i(\circt_i; \circt_i) + \int_{\circt_i}^{t_i} \by^{(i)}(s_i)\cdot ds_i \geq V_i(\emptyset; t_i).
\end{displaymath}
Maximizing revenue corresponds to maximizing the expected sum of the interim
payments $q_i(t_i)$ and thus of minimizing $V_i(\circt_i; \circt_i)$. Since
$\forall t_i, V_i(\emptyset; t_i) \geq -\sum_{j \in N \sm i} \eta_{i \gets
j}$, we can maximize the feasible region for IR payments by setting
$V_i(\emptyset; t_i) = -\sum_{j \in N \sm i} \eta_{i \gets j}$ with an outside
option that allocates to all $j \in N \sm i$ when $i$ does not participate.
That is, we set $x_j(\hat{t}_i = \emptyset, \hat{\bt}_{-i}) = \ind{i \neq j}$
for all $i,j\in N$ and $\bt_{-i} \in T_{-i}$.

\paragraph{Sufficiency of checking interim IR for type $\circt_i$.}
If the interim IR constraint holds for all types $t_i$, then it clearly holds for the type $\circt_i$. Now suppose that $V_i(\circt_i;\circt_i) \geq V_i(\emptyset; \circt_i)$. Note that given the optimal outside option of allocating to all remaining bidders, we have that for every $t_i \in T_i$, 
\begin{align}\label{eq:optimal_outside_rm1}
V_i(\emptyset; t_i) = -\sum_{j \in N \sm i} \eta_{i \gets j}.
\end{align}

Then for every $t_i \in T_i$
\begin{align*}
V_i(t_i; t_i) &= V_i(\circt_i;\circt_i) + \int_{\circt_i}^{t_i} \by^i(s_i)\cdot ds_i
\\ & \geq V_i(\circt_i; \circt_i) + \sum_{j \in N \sm i} (-\eta_{i \gets j} - (-\ubar{\eta}_{i \gets j}))
\\ &= V_i(\circt_i; \circt_i) + V_i(\emptyset; t_i) - V_i(\emptyset; \circt_i)
\\ &\geq V_i(\emptyset; t_i)
\end{align*}
where the first inequality uses that $t_{i,i} = v_i \geq \ubar{v}_i$, $t_{i,j} = -\eta_{i \gets j} \leq -\ubar{\eta}_{i \gets j}$ and $\boldsymbol{0}\leq\by^i\leq 1$ as an allocation vector, the second equality follows from \eqref{eq:optimal_outside_rm1}, and the last inequality follows from our assumption that $V_i(\circt_i; \circt_i) - V_i(\emptyset; \circt_i) \geq 0$. 

\paragraph{Optimal IR Constant.} Since it suffices to satisfy interim IR for the critical type $\circt_i$, for each $i \in N$, in order to maximize revenue we make the IR constraint bind at $\circt_i$
\begin{align}
    V_i(\circt_i, \circt_i) = V_i(\emptyset, \circt_i) = -\sum_{j \in N \sm i} \ubar{\eta}_{i \gets j},
\end{align}
where we plugged in the utility of non-participation of bidder $i$ under the optimal outside option allocation. 
\end{proof}

\subsection{Characterizations in Scenario 2: Outgoing Externalities}
\label{appendix:BNIC_char}
Recall that each bidder $i$'s type is the vector of the form $t_i = v_i e_i - \sum_{j \in N \sm i} \eta_{j \gets i} e_j \in \R^n$ where $e_k$ denotes the
$k$th vector of the standard basis. 
Note that in this
scenario, bidder $i$'s expected outside option utility $V_i(\emptyset; t_i)$
does not depend on $t_i$.

For each bidder $i \in N$ with true type $t_i$ and bid $\hat{t}_i$ the interim expected allocation to bidder $i$ is given by $\cex{x_i(\hat{t}_i, \bt_{-i})}{t_i}=\E_{\bt_{-i}}[x_i(\hat{t}_i, \bt_{-i})]$ and the interim expected payment for bidder $i$ is $\cex{p_i(\hat{t}_i, \bt_{-i})}{t_i} = \E_{\bt_{-i}}[p_i(\hat{t}_i, \bt_{-i})]$. Again, under the given assumption of independent bidder types, the interim functions do not depend on bidder $i$'s true type $t_i$, so we define $y_i(\hat{t}_i)\eqdef \E[x_i(\hat{t}_i, \bt_{-i})]$ to be the interim expected allocation function for bidder $i$ and likewise, $q_i(\hat{t}_i)\eqdef\E[p_i(\hat{t}_i, \bt_{-i})]$ to be the interim expected payment function. The following two propositions then characterize BNIC mechanisms, and interim IR for BNIC mechanisms.  
\begin{proposition}\label{prop:BNIC_char}%
	Assume that private types are of the form $t_i  =v_i e_i - \sum_{j \in N \sm i} \eta_{j \gets i} e_j $ for
	each bidder $i \in N$. The mechanism $(\bx, \bp)$ is BNIC if and only if for each
	bidder $i\in N$:
\begin{enumerate}[label=(\roman*)]
	\item\label{it:non-decreasing} there exists a non-decreasing function
		$\ty_i:[\ubar v_i, \bar v_i]\to[0,1]$ such that the interim allocation
		satisfies $y_i(v_i, \be_{\gets i})=\ty_i(v_i)$ for all but countably
		many $v_i$ and all
	$\be_{\gets i}\in\prod_{j\in N\sm
	i}[\ubar\eta_{j\gets i},\bar\eta_{j\gets i}]$.\label{eq:BNIC_allocchar}
\item\label{it:payment} the interim payment function is given by
\begin{equation} \label{eq:prop3_payment_app}
q_i(t_i) = v_i\cdot \ty_i(v_i) - \int_{\ubar{v}_i}^{v_i} \ty_i(v) dv -
\sum_{j \in N \sm i}\cex*{\eta_{i \gets j}\cdot x_j(t_i,  \bt_{-i})}{t_i}
- C_i 
\,,
\end{equation}
where $C_i$ is an arbitrary integration constant. 
\end{enumerate}
Furthermore, if these conditions are satisfied, then $V_i(t_i; t_i)$ is constant with respect to $\be_{\gets i}$ for almost every $v_i \in [\ubar{v}_i, \bar{v}_i]$, and $C_i = V_i(\ubar{v}_i e_i - \sum_{j \in N \sm i} \eta_{j \gets i} e_j )$ for all $\be_{\gets i} \in \prod_{j \in N \sm i} [\ubar{\eta}_{j \gets i}, \bar{\eta}_{j \gets i}]$.
\end{proposition}
\begin{proof}{}
We first show the necessary implications of IC.
Writing Definition \ref{def:bnic} for $t_i = v_i e_i - \sum_{j \in N \sm i} \eta_{j \gets i} e_j$ as the true type and $\hat{t}_i = \hat{v}_i e_i - \sum_{j \in N \sm i} \hat{\eta}_{j \gets i} e_j $ as the reported type, and then vise versa, i.e., 
\begin{align} 
& V_i(t_i; t_i) \geq V_i(\hat{t}_i; t_i) \label{eq:temp1}
\\ & V_i(\hat{t}_i; \hat{t}_i) \geq V_i(t_i; \hat{t}_i). \notag 
\end{align}
Combining the two inequalities yields 
\begin{align*}
y_i(t_i)(v_i- \hat{v}_i) \geq y_i(\hat{t}_i)(v_i - \hat{v}_i). 
\end{align*}
By Lemma \ref{lem:monotone} below, this implies condition
\ref{it:non-decreasing}, so we let $\tilde{y}_i(v_i)\eqdef y_i(v_i e_i - \sum_{j \in N \sm i} \eta_{j \gets i} e_j )$ for any $\be_{\gets i}$. 

Note also that the inequality \eqref{eq:temp1}, by adding and subtracting the term $\hat{v}_iy_i(\hat{t}_i)$ to the right hand side and regrouping terms, can be written equivalently as
\begin{align} \label{eq:prop3_IC2}
& \forall t_i, \hat{t}_i \in T_i, \notag
\\ & V_i(t_i; t_i) \geq V_i(\hat{t}_i; \hat{t}_i) + (v_i - \hat{v}_i)y_i(\hat{t}_i).
\end{align}

Then plugging in $t_i = v_i e_i - \sum_{j \in N \sm i} \eta_{j \gets i} e_j$
and $\hat{t}_i = v_i e_i - \sum_{j \in N \sm i} \hat{\eta}_{j \gets i} e_j$
into the preceding inequality yields $V_i(t_i) \geq V_i(\hat{t}_i)$, where we
recall that $V_i(t_i) \eqdef V_i(t_i; t_i)$. Swapping the roles of $t_i$ and $\hat{t}_i$ yields the inequality in the opposite direction, and we have that $V_i(t_i)$ is independent of $\be_{\gets i}$:
\begin{align*}
\forall v_i, \forall \be_{\gets i}, \forall \hat{\be}_{\gets i},
\ V_i\lrp*{v_i e_i - \sum_{j \in N \sm i} \eta_{j \gets i} e_j } = V_i\lrp*{v_i e_i - \sum_{j \in N \sm i} \hat{\eta}_{j \gets i} e_j }.
\end{align*}
We henceforth define $\tilde{V}_i(v_i) \eqdef V_i(v_i e_i - \sum_{j \in N \sm i} \eta_{j \gets i} e_j )$ for any $\be_{\gets i}$. 

To prove \ref{it:payment}, we first note that $V_i(t_i)$ is convex in $v_i$. \eqref{eq:temp1} implies that 
\begin{align*}
V_i(t_i, t_i) = \max_{\hat{t}_i \in T_i} y_i(\hat{t}_i)v_i - 
\sum_{j \in N \sm i}\cex*{\eta_{i \gets j}\cdot x_j(\hat{t}_i,  \bt_{-i})}{t_i} - q_i(\hat{t}_i).
\end{align*}
Thus, $V_i(t_i)$ is the maximum of a family of linear functions of $v_i$ and is thus convex in $v_i$. \eqref{eq:prop3_IC2} implies that $y_i(t_i)=\tilde{y}_i(v_i)$ is a subderivative of $V_i(t_i) = \tilde{V}_i(v_i)$. In fact, since $\tilde{V}_i$ is convex in $v_i$, it is differentiable almost everywhere and 
\begin{align*}
y_i(t_i) = \tilde{y}_i(v_i) = \frac{\partial \tilde{V}_i(v_i)}{\partial v_i} \text{a.e.}
\end{align*}
Further, this implies that
\begin{align} \label{eq:prop3_U}
\tilde{V}_i(v_i)= \int_{\ubar{v}_i}^{v_i} \tilde{y}_i(v)dv + \tilde{V}_i(\ubar{v}_i)
\end{align}

Now plugging in the following expression for $V_i$,
\begin{displaymath}
V_i(t_i) = v_i \cdot y_i(t_i) - 
\sum_{j \in N \sm i}\cex*{\eta_{i \gets j}\cdot x_j(t_i,  \bt_{-i})}{t_i}
-q_i(\hat t_i)
\,.
\end{displaymath}
and solving for $q_i(t_i)$, we get 
\begin{align}
q_i(t_i) = v_i \tilde{y}_i(v_i) - \int_{\ubar{v}_i}^{v_i} \tilde{y}_i(v) dv -
\sum_{j \in N \sm i}\cex*{\eta_{i \gets j}\cdot x_j(t_i,  \bt_{-i})}{t_i} - \tilde{V}_i(\ubar{v}_i).
\end{align}

We next show the sufficiency of the conditions for IC, by proving the equivalent condition for IC, \eqref{eq:prop3_IC2}. We have that $\forall i \in N, \forall t_i =v_i e_i - \sum_{j \in N \sm i} \eta_{j \gets i} e_j , \forall \hat{t}_i =\hat v_i e_i - \sum_{j \in N \sm i} \hat \eta_{j \gets i} e_j ,$
\begin{align*}
V_i(t_i) - V_i(\hat{t}_i) & = \int_{\hat{v}_i}^{v_i} \tilde{y}_i(v)dv 
\\  &\geq \tilde{y}_i(\hat{v}_i) (v_i - \hat{v}_i)
\end{align*}
where the first equality follows from \eqref{eq:prop3_U} and the inequality follows from condition~\ref{it:non-decreasing} that $\tilde{y}_i(v_i)$ is increasing in $v_i$. 

\begin{lemma}\label{lem:monotone}
For $d\geq 1$, let $f: \R\times\R^d \rightarrow \R$ be
a bounded function such that
\begin{displaymath}
	\label{eq:sw2_lem1}
f(x_2, y_2) (x_2 - x_1) \geq f(x_1, y_1)(x_2 - x_1),
\quad (x_1,y_1), (x_2, y_2)\in\R\times\R^d
\,.
\end{displaymath}
Then, there exists a non-decreasing function $\tilde f:\R\to\R$ such that
$f(x, y)= \tilde f(x)$ for all $y\in \R^d$ and for all but at most countably
many values of $x\in\R$.
\end{lemma}

	By a rescaling and shifting of $f$ we assume without loss of generality
	that the range of $f$ is contained in $[0,1]$. Let us now define $S\eqdef
\set{x \in \R: \exists y_1, y_2 \in \R^d \st f(x, y_2) \neq f(x, y_1)}$ and for
$n\geq 1$, $S_n\eqdef\set{x \in \R: \exists y_1, y_2 \in \R^d \st f(x, y_2)
- f(x,y_1) \geq 1/n}$ and observe that $S = \cup_{n\geq 1}S_n$.

We now prove that $|S_n|\leq n$. Indeed, consider $x_1<\dots< x_m$, $m$ distinct
points in $S_n$, and for each $k\in[m]$, $y_k^1, y_k^2$ such that $f(x_k, y_k^2)
- f(x_k, y_k^1)\geq 1/n$. Then
\begin{align*}
	\frac{m}{n} 
	& \leq  \sum_{k = 1}^m \big[f(x_k, y_k^2) - f(x_k, y_k^1)\big]\\
& = f(x_m, y_m^2) - \sum_{k=2}^m [f(x_k,  y_k^1) - f(x_{k-1}, y_{k-1}^2)]
- f(x_1, y_1^1)\\
& \leq f(x_m, y_m^2) - f(x_1, y_1^1) \leq 1
\,,
\end{align*}
where the first inequality uses the definition of $S_n$, the equality is
summation by parts, the second inequality uses our assumption on $f$ and the
last inequality uses that the range of $f$ is contained in $[0,1]$. It then
follows that $m\leq n$, i.e. that $|S_n|\leq n$, which in turn implies that $S$
is countable.

Define $\tilde f$ by $\tilde f(x) = f(x, y)$ for $x\notin S$ (this definition
does not depend on the choice of $y$ by definition of $S$). Then our assumption
on $f$ immediately implies that $\tilde f$ is non-decreasing on $\R\sm S$. We
can thus extend $\tilde f$ to a non-decreasing function defined over all of
$\R$ (for example by right continuity). The resulting $\tilde f$ satisfies
the stated requirements.
\end{proof}

\begin{proposition} \label{prop:IRchar}
	Suppose private types are of the form $t_i =  v_i e_i - \sum_{j \in N \sm i} \eta_{j \gets i} e_j \in T_i$ for each
	bidder $ i \in N$. Then a BNIC mechanism satisfies the interim IR
	constraint $V_i(t_i; t_i)\geq V_i(\emptyset; t_i)$ for all $t_i\in T_i$, if and only if this
	condition is satisfied for some type of the form $ \ubar{v}_i e_i - \sum_{j \in N \sm i} \eta_{j \gets i} e_j $, where $\eta_{j \gets i} \in [\ubar{\eta}_{j \gets i}, \bar{\eta}_{j \gets i}]$ for $j \in N \sm i$.
\end{proposition}
\begin{proof}{}
Note that BNIC implies \eqref{eq:prop3_U}, and since $\tilde{y}_i \geq 0$, we have that $V_i(t_i; t_i) \geq \tilde{V}_i(\ubar{v}_i)$ for all $t_i \in T_i$. Since $V_i(\emptyset; t_i)$ is independent with respect to $t_i$, it is both necessary and sufficient for IR to hold that the IR condition holds for some type of the form $ \ubar{v}_i e_i - \sum_{j \in N \sm i} \eta_{j \gets i} e_j $, for each bidder $i \in N$. 
\end{proof}

\section{Welfare Maximization}
\subsection{Proof of Theorem \ref{prop:vcgprops}} \label{appendix:vcgprops}
We show that the specified VCG mechanism (1) satisfies DSIC, (2) ex-post IR,
and (3) uses nonnegative payments. Recall that in this setting, private types
are of the form $t_i = v_i e_i - \sum_{j \in N \sm i} \eta_{i \gets j}$, for $i \in N$. 
\begin{enumerate}
    \item For all  $i \in N$ and all $t_i, \hat{t}_i \in \Theta_i, \bt_{-i}, \hat{\bt}_{-i} \in \Theta_{-i}$, let us temporarily define the following quantities for ease of notation. Note the only quantity varying in the following terms is bidder $i$'s bid, while all other parameters are fixed.
    \begin{align*}
    & x_i \eqdef x_i(t_i, \hat{\bt}_{-i}), \hat{x}_i \eqdef x_i(\hat{t}_i, \hat{\bt}_{-i})
    \\ & x_j \eqdef x_j(t_i, \hat{\bt}_{-i}), \hat{x}_j \eqdef x_j(\hat{t}_i,
	\hat{\bt}_{-i}), x_j^i \eqdef x_j(t_i = \emptyset, \bt_{-i}) , \, \text{ for } j \in N \sm i
    \\ & p_i \eqdef p_i(t_i, \hat{\bt}_{-i}), \hat{p}_i \eqdef p_i(\hat{t}_i, \hat{\bt}_{-i})
    \\ & W_i \eqdef v_i - \sum_{j \in N \sm i} \eta_{j\gets i}, \hat{W}_i
	\eqdef \hat{v}_i - \sum_{j \in N \sm i} \eta_{j\gets i}
    \\ & W_j = v_j - \sum_{k \in N \sm j} \eta_{k\gets j}, \hat{W}_j \eqdef v_j - \sum_{k \in N \sm \set{j, i}} \eta_{k\gets j} - \hat{\eta}_{i \gets j}\text{, for } j \in N \sm i
	\end{align*}     
	We show that the following expression is nonnegative, which is precisely the statement of DSIC:
    \begin{align*}
    &u_i\of[\big]{\bx(t_i, \hat\bt_{-i}), p_i(t_i,\hat\bt_{-i}); \bt}
- u_i\of[\big]{\bx(\hat\bt), p_i(\hat\bt); \bt}
    \\ & = (x_i - \hat{x}_i)v_i - \sum_{j \in N \sm i} \eta_{i \gets j} (x_j - \hat{x}_j) 
    \\ & \quad \quad \quad \quad \quad \quad + \sum_{j \in N \sm i} (-W_j^i(x_j^i - x_j) - \eta_{j\gets i} x_i + W_j^i(x_j^i - \hat{x}_j) - \eta_{j\gets i} \hat{x}_i)
    \\ & = (x_i - \hat{x}_i)(v_i - \sum_{j \in N \sm i} \eta_{j\gets i}) + \sum_{j \in N \sm i} (W_j^i - \eta_{i \gets j})(x_j - \hat{x}_j)
    \\ & = (\ind{W_i \geq 0} - \ind{\hat{W}_i \geq 0}) W_i + \sum_{j \in N \sm i} W_j (\ind{W_j \geq 0} - \ind{\hat{W}_j \geq 0})
    \\ & \geq 0.
    \end{align*} 
For the first equality we used the second expression of the payment rule in \eqref{eq:sw1_p}, we regrouped terms and used the definitions of $W_i, W_j$ for the second and third equalities. The final inequality holds because 
    \begin{align*}
    \ind{W_i \geq 0} - \ind{\hat{W}_i \geq 0} =  
    \begin{cases}
	1 & \text{if $W_i\geq 0$ and $\hat{W}_i < 0$} \\
	-1 & \text{if $W_i < 0$ and $\hat{W}_i \geq 0$} \\
	0 & \text{otherwise,}
    \end{cases}
    \end{align*} 
    and likewise for $\ind{W_j \geq 0} - \ind{\hat{W}_j \geq 0}$, implying that each term in the summation is nonnegative. 
    \item Let $\bt$ be an arbitrary type realization. Showing ex-post IR is equivalent to showing 
    \begin{align*}
	& v_i x_i - \sum_{j \in N \sm i} \eta_{i \gets j}x_j - p_i \geq u_i(\bx(\emptyset, \bt_{-i}), p_i(\emptyset, \bt_{-i}); \bt) = -\sum_{j \in N \sm i} \eta_{i \gets j} x_j^i.
    \end{align*}
    Plugging in the payment rule, we get the equivalent inequality
    \begin{align*}
    v_i x_i - \sum_{j \in N \sm i} \left( W_j^i(x_j^i - x_j) + \eta_{j\gets i}x_i \right) \geq -\sum_{j \in N \sm i} \eta_{i \gets j}(x_j^i - x_j)
    \end{align*}
    Rearranging and regrouping terms, we get that this is equivalent to 
    \begin{align*}
    W_i x_i - \sum_{j \in N \sm i} W_j(x_j^i - x_j) \geq 0
	\end{align*}        
	Since $x_i = \ind{W_i \geq 0}$, the first term is always nonnegative. The terms in the summation are likewise nonnegative since 
	\begin{align*}
    x_j^i - x_j = 
    \begin{cases}
		+1 & \text{if $W_j\geq 0$ and $W_j^i < 0$} \\
		-1 & \text{if $W_j < 0$ and $W_j^i \geq 0$} \\
		0 & \text{otherwise}.
    \end{cases}
    \end{align*}   
    Thus, the IR constraint is satisfied for all types $\bt$.
    \item Let $\bt$ be an arbitrary type realization. Note that since $\eta_{j\gets i} \geq 0$, we can lower bound the payments in \eqref{eq:sw1_p} by 
    \begin{align*}
	p_i \geq \sum_{j \in N \sm i} W_j^i\prn[\big]{\ind{W_j^i \geq 0} - \ind{W_j\geq 0}}.
    \end{align*}
    We have that
    \begin{align*}
    \ind{W_j^i \geq 0} - \ind{W_j\geq 0} = 
    \begin{cases}
		+1 &\text{if $W_j^i \geq 0$ and $W_j<0$} \\
		-1 &\text{if $W_j^i < 0$ and $W_j\geq 0$} \\
		0 &\text{otherwise.}
    \end{cases}
    \end{align*}
    Matching up the cases, we get that $p_i \geq 0$, so payments are nonnegative. 
\end{enumerate}

\subsection{Proof of Proposition \ref{prop:sw2opt-impossibility}} \label{appendix:sw2_impossibility}
Consider a uniform type distribution $F$ over two points: $\bt^a = (\hat{v}_1, \be_{1 \to}^a, \hat{\bt}_{-1})$ and $\bt^b = (\hat{v}_1, \be_{1 \to}^b, \hat{\bt}_{-1})$, where $\hat{v}_1 \in \R_{+}$ and $\hat{\bt}_{-1} = (t_j)_{j \neq 1}$ are arbitrary, fixed values. We choose $\be_{1 \to}^a = (\eta_{j \gets 1})_{j \neq 1}$ and $\be_{1 \to}^b = (\eta_{j \gets 1})_{j \neq 1}$ satisfying
\begin{equation}\label{eq:ineq}
	\sum_{j \in N \sm 1} \eta_{j \gets 1}^a < \hat{v}_1
	< \sum_{j \in N \sm 1} \eta_{j \gets 1}^b.
\end{equation} 
For instance, we can take $\be_{1 \to}^a = \boldsymbol{0}$ and $\be_{1 \to}^b = 2\hat{v}_1 \boldsymbol{1}$. 

Let us consider an allocation $\bx(\cdot) \in \mathcal{X}_{BNIC}(F)$. By
\cref{prop:BNIC_char} there exists a non-decreasing function $\tilde{y}$ such
that $\forall \bt \in T$, $x_1(\bt) = \E[x_1(\bt)|t_1] = \tilde{y}(v_1)$.
However, under the distribution $F$, $v_i$ only takes the single value
$\hat{v}_i$, so the function $x_1(\bt)$ must be constant-valued. We now
consider two cases depending on the sign of this constant value.

If $x_1(\bt) > 0$ for all $\bt\in T$, then define $\by(\cdot)$ by $y_1(\bt) = 0$ and $y_j(\bt) = x_j(\bt)$ for all $j \neq 1$. We have that 
\begin{align*}
SW(x; \bt^b) - SW(y; \bt^b) = \prn[\Big]{\hat{v}_1 - \sum_{j \in N \sm i} \eta_{j \gets 1}^b} \prn[\big]{(x_1(\bt^b) - y_1(\bt^b)} < 0
\end{align*}
where the strict inequality follows from \eqref{eq:ineq} and that $\forall \bt, x_1(\bt) > 0 = y_1(\bt)$.

Likewise, if $x_1(\bt) \leq 0$ for all $\bt\in T$, define $\by(\cdot)$ by $y_1(\bt) = 1$ and $y_j(\bt)=x_j(\bt)$ for all $j \neq 1$. Then 
\begin{align*}
SW(x; \bt^a) - SW(y; \bt^a)
= \prn[\Big]{\hat{v}_1 - \sum_{j \in N \sm 1} \eta_{j \gets 1}^a}\prn[\big]{x_1(\bt^a) - y_1(\bt^a)} < 0
\end{align*}
where the strict inequality follows from \eqref{eq:ineq} and that $\forall \bt, x_1(\bt) \leq 0 < 1 = y_1(\bt)$. 

Thus, we have shown that for any BNIC allocation, there exists some type realization for which a different BNIC allocation rule yields strictly greater social welfare. This is precisely the statement in \eqref{eq:sw_ineq}.

\subsection{Proof of Theorem \ref{prop:sw2_alloc}} \label{appendix:sw2_alloc_proof}
\paragraph{Allocations.} To solve for the form of the expected welfare maximizing allocation function
satisfying the IC constraints, we first express the objective in terms of the
interim allocations $y_i(t_i)$. Defining $W_i(t_i)\eqdef
v_i - \sum_{j \in N \sm i} \eta_{j\gets i}$ as in Section \ref{sec:sw1}, we have 
\begin{displaymath}
\ex[\big]{\sw(\bx; \bt)}
= \sum_{i \in N} \ex{W_i(t_i) x_i(\bt)}
= \sum_{i \in N} \ex[\big]{\cex{W_i(t_i) x_i(t_i, \bt_{-i})}{t_i}}
= \sum_{i \in N} \ex{W_i(t_i) y_i(t_i)} \,.
\end{displaymath}
Now, the BNIC characterization from Proposition~\ref{prop:BNIC_char} implies
that there must exist functions $\ty_i:[\ubar v_i, \bar v_i]\to[0,1]$ such that $y_i(t_i) = \ty_i(v_i)$ for almost all $t_i\in T_i$. Plugging in this
representation above, we get 
\begin{displaymath}
	\ex{\sw(\bx; \bt)}
	= \sum_{i \in N} \ex{W_i(t_i) \ty_i(v_i)}
=  \sum_{i \in N} \ex[\big]{\ty_i(v_i) \cex{W_i(t_i)}{v_i}} \,.
\end{displaymath}
Noting the linearity of the objective in $\ty_i$, we find that the
optimal allocation rule is 
\begin{displaymath} 
	\ty_i(v_i)
= \ind[\big]{\cex{W_i(t_i)}{v_i} \geq 0}
 = \ind[\bigg]{v_i - \sum_{j \in N \sm i} \cex{\eta_{j \gets i}}{v_i} \geq 0}\,.
\end{displaymath}
Under the given assumptions, $\ty_i(v_i)$ is non-decreasing in $v_i$, so BNIC is satisfied. Finally, note that since we can express the objective function and
constraints only in terms of the interim allocations $y_i$ for $i \in N$, we can without loss of
generality for all $i \in N$, set the allocation rule $x_i(\bt) = y_i(t_i)=\ty_i(v_i)$.

\paragraph{Payments.} The induced interim payment rule $q_i(t_i) = \cex{p_i(\bt)}{t_i}$ associated with
$y_i$ as derived in Proposition~\ref{prop:BNIC_char}, condition (ii) is
\begin{align}\label{eq:sw2_pay1}
q_i(t_i) &= v_i\cdot y_i(t_i) - \int_{\ubar{v}_i}^{v_i} \ty_i(v) dv
- \sum_{j \in N \sm i} \cex[\big]{\eta_{i \gets j} x_j(t_i,  t_{-i})}{t_i}
- C_i  
\,.
\end{align}
Recall that the constant term $C_i$ is set
such that the payment function satisfies IR. By Proposition \ref{prop:IRchar}, it
suffices to check IR for any type of the form $t_i = (\ubar{v}_i,
\be_{\gets i})$, for each $ i \in N$. Here, bidder
$i$'s expected utility $V_i(\emptyset; t_i)$ if she doesn't participate is the sum of the
externalities effects from the allocations $x_j(t_i = \emptyset, \bt_{-i})$ in the welfare-maximizing
auction run with the remaining set $N \sm i$ of bidders, and given the assumption of  $v_j \mapsto v_j - \sum_{k \in N \sm \set{j, i}} \E[\eta_{k\gets j}|v_j]$ non-decreasing, we have
\begin{align} \label{eq:sw2_outsideutil}
	V_i(\emptyset; t_i) &= \sum_{j \in N \sm i}\ex*{\eta_{i \gets j} x_j(t_i = \emptyset, \bt_{-i})(\bt_{-i})} \nonumber
\\&=  \quad \sum_{j \in N \sm i} \ex[\Bigg]{\eta_{i \gets j}
\ind[\bigg]{v_j \geq \sum_{\mathclap{k \in N \sm \set{{j,i}}}} \E[\eta_{k\gets j}|v_j]}} 
\end{align}
Then any payment rule of the form \eqref{eq:sw2_pay1} with the constant $C_i$ set greater than or equal to $V_i(\emptyset; t_i)$ in \eqref{eq:sw2_outsideutil} will give us an IR mechanism. 

Finally, since the objective function and constraints can be expressed solely
in terms of the interim payments $q_i$, we can set $p_i(\bt)\eqdef q_i(t_i)$.
Under the given assumption that $v_i \mapsto v_i - \sum_{j \in N \sm i} \E[\eta_{j \gets i} | v_i]$ is non-decreasing, we can re-express the allocation rule as 
\begin{equation}\label{eq:sw2_x2}
    x_i(\bt) = y_i(t_i)
	= \ind[\bigg]{v_i \geq \sum_{\mathclap{j \in N \sm i}}\cex{\eta_{j \gets i}}{v_i}}
	= \ind[]{v_i \geq \tau_i}
\end{equation} 
The integral term in \eqref{eq:sw2_pay1} then becomes
\begin{align*}
\int_{\ubar{v}_i}^{v_i} \ind[]{v \geq \tau_i}  dv 
&=(v_i - \tau_i)
\cdot\ind[]{v_i \geq \tau_i}
= (v_i - \tau_i) \cdot y_i(t_i)
\end{align*}
Plugging in the above expression, along with the form of the allocation \eqref{eq:sw2_x2}, into \eqref{eq:sw2_pay1} completes the proof.

\section{Revenue Maximization}

\subsection{Proof of Proposition~\ref{prop:impossibility}}\label{app:imposs}

An instance of the $n$-item auction is a distribution $F$ over $n$-tuples of item valuations. We make the standard assumption that item valuations are non-negative (i.e.\ the marginals of $F$ are supported on $\R_{+}$), and denote by $\mathcal{V}$ the support of $F$.
Let $s:\R^n\to\R^n$ be the involution that flips the signs of all but the first of its input's coordinates: $s(v) = (v_1, -v_2,\dots,-v_n)$.
We define the distribution $\widetilde F = s_*(F)\otimes\delta_0\otimes\cdots\otimes\delta_0$ over type profiles in the digital good auction, where $s_*(F)$ is the pushforward of $F$ through $s$ and $\delta_0$ denotes the Dirac measure at $0\in\R^n$.
In other words, the support of $\widetilde F$ is $T=s(\mathcal{V})\times\set{0}\times\cdots\times\set{0}$, and $\bt$ is distributed according to $\widetilde F$ iff $s^{-1}(t_1)=s(t_1)$ is distributed according to $F$ and $t_2,\dots, t_n$ are constant equal to $0\in\R^n$. 

Let $(x,p)$ be a mechanism for the $n$-item auction, with $x:\mathcal{V}\to[0,1]^n$ describing the allocation probabilities of each of the $n$ items, and $p:\mathcal{V}\to\R_{+}$ being the corresponding payment function. For each $(x,p)$, we define a corresponding mechanism $(\tilde x, \tilde p)$ for the $n$-player digital good auction as follows: for each $t\in T$,
\begin{align}\label{eq:reduction-map}
	\begin{cases}
	\tilde x_1(t) = x_1\big(s(t_1)\big)\\
	\tilde p_1(t) = p\big(s(t_1)\big)
\end{cases}
\text{and for $2\leq i\leq n$,}\quad
\begin{cases}
	\tilde x_i(t) = 1-x_i\big(s(t_1)\big)\\
	\tilde p_i(t) = 0
\end{cases}.
\end{align}
Then, 
\begin{equation}\label{eq:ic-reduction}
	\begin{split}
	&\forall (v,v')\in\mathcal{V}^2,\;\ip*{v}{x(v)} - p(v)
	\geq \ip*{v}{x(v')} - p(v')\\
	\iff&\forall (t_1, t_1')\in T_1^2,\; \ip*{s(t_1)}{x\big(s(t_1)\big)} - p\big(s(t_1)\big)
	\geq \ip*{s(t_1)}{x\big(s(t_1')\big)} - p\big(s(t_1')\big)\\
	\iff&\forall (t_1, t_1')\in T_1^2,\;
	\begin{aligned}[t]
	&\,t_{11}\tilde x_1(t_1, 0) -\sum_{i=2}^n t_{1i}\big(1-\tilde x_i(t_1,0)\big)
	- \tilde p_1(t_1, 0)\\
	\geq &\,t_{11}\tilde x_1(t_1', 0) -\sum_{i=2}^n
	t_{1i}\big(1-\tilde x_i(t_1',0)\big) - \tilde p_1(t_1', 0)
	\end{aligned}\\
	\iff&\forall (t_1, t_1')\in T_1^2,\;
	\ip*{t_{1}}{\tilde x(t_1,0)} - \tilde p_1(t_1, 0)
	\geq \ip*{t_{1}}{\tilde x(t_1',0)} - \tilde p_1(t_1', 0).
\end{split}
\end{equation}
We used that $s$ induces a bijection from $T_1\eqdef s(\mathcal{V})$ to $\mathcal{V}$ in the first equivalence, and the definitions of $s$, $\tilde x$ and $\tilde p$ in the second.
Adding $\sum_{i=2}^n t_{1i}$ to both sides of the inequality yields the third equivalence. Recall that under $\widetilde F$, each of the type vectors $t_2,\dots, t_n$ is supported on $0\in\R^n$ a.s., hence truthful reporting of players $2$ to $n$ is trivially guaranteed.
Finally,~\eqref{eq:ic-reduction} shows that $(x,p)$ is IC iff $(\tilde x,\tilde p)$ satisfies the IC constraint of player $1$.

We now look at the participation constraints. In the $n$-item auction, non-participation implies that the buyer is allocated no item (and no payment is collected), resulting in a vanishing reservation utility.
In the $n$-player digital good auction, we set the allocation of player $1$ in case of non-participation to be $\tilde x_i(\emptyset, 0) = \ind{i\neq 1}$, which results in the worst possible reservation utility $\sum_{i=2}^n t_{1i}$ for type $t_1$.
Then, following the same steps as in~\eqref{eq:ic-reduction}:
\begin{align*}
	&\forall v\in\mathcal{V},\;\ip*{v}{x(v)} - p(v)
	\geq 0\\
	\iff&\forall t_1\in T_1,\; \ip*{s(t_1)}{x\big(s(t_1)\big)} - p\big(s(t_1)\big)
	\geq 0\\
	\iff&\forall t_1\in T_1,\;
	\,t_{11}\tilde x_1(t_1, 0) -\sum_{i=2}^n t_{1i}\big(1-\tilde x_i(t_1,0)\big)
	- \tilde p_1(t_1, 0) \geq 0\\
	\iff&\forall t_1\in T_1,\;
	\ip*{t_{1}}{\tilde x(t_1,0)} - \tilde p_1(t_1, 0)
	\geq \sum_{i=2}^n t_{1i}.
\end{align*}
This shows that $(\tilde x, \tilde p)$ guarantees participation of player 1 iff $(x,p)$ is IR\@. Furthermore, the participation constraints of player $2$ to $n$ are trivially satisfied under $\widetilde F$. Note also that these constraints require the payment functions $\tilde p_i$ to be identically zero for $2\leq i\leq n$.

The previous two paragraphs show that there exists a mapping
$M:(x,p)\mapsto(\tilde x, \tilde p)$ between mechanisms, namely the one given
by \eqref{eq:reduction-map}, that establishes a one-to-one correspondence
between (i) IC and IR mechanisms for the $n$-item auction under distribution
$F$, and (ii) IC and IR mechanisms for the $n$-player digital good auction
under distribution $\widetilde F$ (with outside option defined as above).
Furthermore, the expected revenue of each mechanism $(x,p)$ is equal to the
expected revenue of the corresponding mechanism $(\tilde x,\tilde p) = M(x,p)$.
Since the outside option we considered makes the IR constraint of player $1$
the slackest, this implies that the range of $M$ contains an optimal mechanism
for the digital good auction. The preimage of such a mechanism under $M$ is
therefore optimal for the $n$-item auction under $F$.

\subsection{Proofs of Theorem~\ref{prop:NEWrm2_BNIC} and Corollary~\ref{cor:restricted-dep-cor}} \label{appendix:rm2_opt}
Recall that for either scenario of private types, for each bidder $i \in N$, with true type $t_i$ and bid $\hat{t}_i$, the overall interim expected allocation is defined as $\cex{\bx(\hat{t}_i, \bt_{-i})}{t_i} = \E_{\bt_{-i}}[x(\hat{t}_i, \bt_{-i})]$ and the interim expected payment as $\cex{p_i(\hat{t}_i,\bt_{-i})}{t_i} = \E_{\bt_{-i}}[p_i(\hat{t}_i, \bt_{-i})]$.
Note that, under the given assumption of independent bidder types, the interim expected allocation and payment functions do not depend on bidder $i$'s true type $t_i$, so we write $\by^{(i)}(\hat{t}_i)\eqdef \E[x(\hat{t}_i, \bt_{-i})]$ to indicate the interim expected allocation function and likewise, $q_i(\hat{t}_i)\eqdef \E[p_i(\hat{t}_i, \bt_{-i})]$ for the interim expected payment.

We begin with the following lemma which provides a characterization of BNIC tailored to restricted-dependency mechanisms. 

\begin{lemma}\label{lem:restricted-BNIC}
	For restricted-dependency mechanisms, the $j$th coordinate of the interim
	allocation vector $\by^{(i)}(t_i)$ only depends on $t_{ij}$, and we write
	$\by^{(i)}(t_i) = \big(y^{(i)}_j(t_{ij})\big)_{j\in N}$. A
	restricted-dependency mechanism is BNIC iff 
\begin{enumerate}[label=(\roman*)]
	\item for each $(i,j)\in N^2$, the function $y^{(i)}_j$ is non-decreasing.
	\item the interim payment of bidder $i\in N$ is given by
		\begin{align}\label{eq:restricted-dep-payment}
			q_i(t_i) = \ip*{t_i}{\by^{(i)}(t_i)}
			- \sum_{j\in N}\int_{\circt_{ij}}^{t_{ij}} y^{(i)}_j(s)ds-C_i,
		\end{align}
		for some constant $C_i$ whose value sets the interim utility of bidder
		$i$ at $\circt_i$.
\end{enumerate}
\end{lemma}

\begin{proof}
	The mechanism is BNIC iff $V_i(t_i; t_i)\geq V_i(t_i'; t_i)$ for each
	$(t_i, t_i')\in T_i^2$. We re-express this condition in terms of interim
	allocations and payments:
	\begin{align*}
		V_i(t_i;t_i)=\ip*{t_i}{\by^{(i)}(t_i)}-q_i(t_i)
		&\geq V_i(t_i';t_i)=\ip*{t_i}{\by^{(i)}(t_i')}-q_i(t_i')\\
		&= \ip*{t_i'}{\by^{(i)}(t_i')}-q_i(t_i')+\ip*{t_i-t_i'}{\by^{(i)}(t_i')}\\
		&=V_i(t_i';t_i')+\ip*{t_i-t_i'}{\by^{(i)}(t_i')}.
	\end{align*}
By a well-known characterization of convexity, this is equivalent to requiring
that the interim utility $t_i\mapsto V_i(t_i;t_i)$ be convex and admits
$\by^{(i)}(t_i)$ as a subgradient for each $t_i\in T_i$. This is in turn
equivalent to $\by^{(i)}$ being a cyclically monotone vector field. Due to the
restricted-dependency assumption, it is easy to see that $\by^{(i)}$ is
cyclically monotone iff $y_j^{(i)}$ is non-decreasing for all $j\in N$. But then,
$y_{j}^{(i)}$ is integrable, hence $\by^{(i)}$ is conservative with (convex)
potential $t_i\mapsto \sum_{j\in N}\int_{\circt_{ij}}^{t_{ij}}
y^{(i)}_j(s)ds$, and we can write
\begin{displaymath}
	V_i(t_i;t_i) -V_i(\circt_i;\circt_i)=\sum_{j\in N}\int_{\circt_{ij}}^{t_{ij}} y^{(i)}_j(s)ds.%
\end{displaymath}
Thus the interim payments take the form 
\begin{align*}
    q_i(t_i) &= \ip*{t_i}{\by^{(i)}(t_i)}-V_i(t_i;t_i) \\
    &= \ip*{t_i}{\by^{(i)}(t_i)}-\sum_{j\in N}\int_{\circt_{ij}}^{t_{ij}}
	y^{(i)}_j(s)ds + V_i(\circt_i;\circt_i) .\qedhere
\end{align*}

\end{proof}

\subsubsection{Proof of Theorem~\ref{prop:NEWrm2_BNIC}}\label{app:rm-alloc}
\paragraph{Deriving the Optimal Allocation.} We first use the form of the interim payment functions from the BNIC characterization of restricted-dependency mechanisms under Scenario 1 given in \cref{lem:restricted-BNIC} to express our objective solely in terms of interim allocation functions $\by^{(i)}$. %
The expected revenue then becomes 
\begin{align*}
\ex*{\sum_{i \in N} p_i(\bt)} &= \ex*{\sum_{i \in N} \ex*{p_i(\bt)|t_i}}
= \sum_{i \in N} \ex*{q_i(t_i)}\\
&= \sum_{i \in N} \ex*{\ip[\big]{\by^{(i)}(t_i)}{t_i} - \sum_{j\in N}\int_{\circt_{ij}}^{t_{ij}} y^{(i)}_j(s_{ij})ds_{ij}} - C_i \\
&= \sum_{i \in N} \ex*{\ip[\big]{\by^{(i)}(t_i)}{t_i} - \sum_{j\in N}\int_{\ubar{t}_{ij}}^{t_{ij}} y^{(i)}_j(s_{ij})ds_{ij}} - C_i' 
\end{align*} 
The first line follows by the law of total expectation, which allows us to express the expected revenue as the sum of interim expected payments. In the last equality, we shifted the lower bounds of integration from components of $\circt_i$ to components of $\ubar{t}_i: = \ubar{v}_i e_i - \sum_{j \in N \sm i} \bar{\eta}_{i \gets j} e_j$ along with the corresponding constant of integration $C_i$ to $C_i'$. The type vector $\ubar{t}_i$ can be considered the ``lowest'' type of bidder $i$, as it yields the lowest valuation on any given allocation over all feasible types, while the critical type $\circt_i$ is the type vector closest to the origin and is the binding type for IR constraints. Originally $C_i = V(\circt_i; \circt_i)$, and now the new constant of integration $C_i'$ sets the value of $V(\ubar{t}_i; \ubar{t}_i)$. The constant term $C_i'$ can be set independently of the allocation functions, and we defer finding the optimal such $C_i'$ (and thus $C_i$) satisfying IR to the last part of this proof, after we have solved for the optimal allocation rules.

Recall that $t_{ii} = v_i, t_{ij} = -\eta_{i \gets j}$, and for streamlined notation, let $f_{ij}$ and$ F_{ij}$ denote the pdf and cdf of $t_{ij}$, respectively, for $i, j \in N$. Note that for $i \neq j$, $f_{ij}(t_{ij}) = f_{\eta_{i \gets j}}(-t_{ij}) $ and $F_{ij}(t_{ij}) = 1-F_{\eta_{i \gets j}}(\eta_{i \gets j})$. 

The expectation of each of the integral summands becomes 
\begin{align*}
\ex*{\int_{\ubar{t}_{ij}}^{t_{ij}} y^{(i)}_j(s_{ij})ds_{ij}} &= \int_{\ubar{t}_{ij}}^{\bar{t}_{ij}} dt_{ij} f_{ij}(t_{ij}) \int_{\ubar{t}_{ij}}^{t_{ij}} y^{(i)}_j(s_{ij}) ds_{ij} \\
&=  \int_{\ubar{t}_{ij}}^{\bar{t}_{ij}}  y^{(i)}_j(s_{ij}) ds_{ij} \int_{s_{ij}}^{\bar{t}_{ij}}f_{ij}(t_{ij}) dt_{ij}\\
&= \int_{\ubar{t}_{ij}}^{\bar{t}_{ij}}  y^{(i)}_j(t_{ij}) (1-F_{ij}(t_{ij})) dt_{ij} \\
&= \ex*{y^{(i)}_j(t_{ij}) \cdot \frac{1-F_{ij}(t_{ij})}{f_{ij}(t_{ij})}}.
\end{align*}
where in the second line, we swapped the order of integration, in the third line, we evaluated the inner integral and renamed the dummy variable $s_{ij}$ to $t_{ij}$, and in the last line,  multiplied by $f_{ij}(t_{ij})/f_{ij}(t_{ij})$ to retrieve the expectation. 

The expected revenue, up to the constant $C_i'$, can be written as
\begin{align*}
&\sum_{i \in N} \ex*{\ip[\big]{\by^{(i)}(t_i)}{t_i} - \int_{\ubar{t}_i}^{t_i} \by^{(i)}(s_{i})\cdot ds_{i}} \\
&= \ex*{\sum_{i \in N}\lrp*{\sum_{j \in N} y^{(i)}_j(t_{ij})t_{ij} - \sum_{j \in N} y^{(i)}_j(t_{ij}) \cdot \frac{1-F_{ij}(t_{ij})}{f_{ij}(t_{ij})}}} \\
&= \ex*{\sum_{i \in N} \cex*{x_i(t)}{t_i} \sum_{j\in N} \underbrace{\lrp*{t_{ij} - \frac{1-F_{ij}(t_{ij})}{f_{ij}(t_{ij})}}}_{=:\phi_{ij}(t_{ij})}} \\
&= \ex*{\sum_{i \in N} x_i(t) \sum_{j \in N} \phi_{ij}(t_{ij})},
\end{align*}
where in the last line we used the tower law of expectation. 

The expected revenue is linear in the allocations $x_i(t)$. Subject to the constraint that $x_i(t) \in [0,1]$ for all $i \in N, t \in T$, the restricted-dependency revenue maximizing allocation can be read off as 
\begin{align*}
x_i(t) = \ind*{\sum_{i \in N} \phi_{ij}(t_{ij}) \geq 0}.
\end{align*}
Since $v_i = t_{ii}$ and $\eta_{i \gets j} = -t_{ij}$, it follows that $\phi_{ii}(t_{ii}) = \phi_{v_i}(v_i)$ and $\phi_{ij}(t_{ij}) = -\phi_{\eta_{i\gets j}}(\eta_{i\gets j})$ and we obtain the expression \eqref{eq:rm1_alloc} for the allocation.

\paragraph{Verifying BNIC.} Note that for $k \in N$, the allocation functions $x_k(\bt)$ are only dependent on $t_{i,k}$, for all $i \in N$.
 Further, $y^{(i)}_k(t_{ik}) = \mathbb{P}[\sum_{j \in N} \phi_{jk}(t_{jk}) \geq 0|t_{ik}]$ is increasing in $t_{i,k}$ since the relevant term $\phi_{ik}(t_{ik})$ is increasing in $t_{ik}$. This can be derived from our assumption that $\phi_{\eta_{i \gets k}}(\eta_{i \gets k})$ is increasing in $\eta_{i \gets k}$, which implies that $-\phi_{\eta_{i \gets j}}(\eta_{i\gets j}) = \phi_{ij}(t_{ij})$ is decreasing in $\eta_{i \gets j}$ and increasing in $t_{ij} = -\eta_{i \gets j}$ . 
 
 Then by \cref{lem:restricted-BNIC}, $y^{(i)}$ and the payment function $p_i(\bt) = q_i(t_i)$ derived from the corresponding interim payment functions in \eqref{eq:restricted-dep-payment} yield a restricted-dependency BNIC mechanism. The constant $C_i$ can be optimized to maximize revenue, as is done in \cref{cor:restricted-dep-cor}, whose proof is given in the following subsection.

\subsubsection{Proof of Corollary~\ref{cor:restricted-dep-cor}}\label{proof:cor:restricted-dep-cor}
Recall that by \cref{prop:IRchar_type1}, the optimal outside option is to allocate $x_j = 1$ to all bidders $j \in N \sm i$ when bidder $i$ does not participate, and the optimal constant $C_i$ in the interim payment \eqref{eq:restricted-dep-payment} is determined by setting the utility of critical type $\circt_i = \ubar{v}_ie_i - \sum_{j \in N \sm i} \ubar{\eta}_{i \gets j} e_j$ equal to its reservation utility:
\begin{align*}
C_i = V_i(\circt_i; \circt_i) =V_i(\emptyset; \circt_i)=  -\sum_{j \in N \sm i} \ubar{\eta}_{i \gets j}.
\end{align*}  

Although taking the payment functions $p_i(\bt)$ to be equal to the interim payments $q_i(t_i)$ given in \eqref{eq:restricted-dep-payment} of the BNIC characterization of \cref{lem:restricted-BNIC} is suitable for maximizing revenue, we can alternatively define the following payment functions in terms of the non-interim allocation functions $x_i(\bt)$:
\begin{align}\label{eq:non-interim-payment1}
p_i(\bt) &= \ip[\big]{\bx(\bt)}{t_i} -\sum_{j \in N} \int_{\circt_{ij}}^{t_{ij}} x_j(s_{ij}, \bt_{-ij}) ds_{ij} - C_i 
\end{align}
where $\bt_{-ij}\in \R^{n^2 -1}$ is the vector collecting all entries of $\bt$ except for $t_{ij}$. 
Note that taking the conditional expectation of this payment respect to $t_i$ yields the correct interim payment form, \eqref{eq:restricted-dep-payment}, of the BNIC characterization of \cref{lem:restricted-BNIC}: 
\begin{align*}
    \cex{p_i(t_i,\bt_{-i})}{t_i}&= \cex*{\ip[\big]{\bx(\bt)}{t_i} }{t_i} -\sum_{j \in N} \cex*{\int_{\circt_{ij}}^{\bar{t}_{ij}} \ind*{s_{ij} \leq t_{ij}} x_j(s_{ij}, \bt_{-ij}) ds_{ij}}{t_i} - C_i \\
    &= \ip[\big]{\cex*{\bx(\bt)}{t_i}}{t_i} - \sum_{j \in N}
	\int_{\circt_{ij}}^{\bar t_{ij}}  \cex*{  \ind*{s_{ij} \leq t_{ij}} x_j(s_{ij}, \bt_{-ij}) }{t_i}ds_{ij}-C_i\\
    &= \ip[\big]{\by^{(i)}(t_i)}{t_i}- \sum_{j \in N} \int_{\circt_{ij}}^{t_{ij}}   y^{(i)}_j(s_{ij}) ds_{ij}-C_i.
\end{align*}
where in the second line we swapped the order of conditional expectation with integration by Fubini's theorem and nonnegativity of the argument, and in the last line recovered the interim allocations.

Given the assumptions that the virtual valuation functions $\phi_{v_i}$ and $\phi_{\eta_{i \gets j}}$ are non-decreasing, we can equivalently express the optimal allocations \eqref{eq:rm1_alloc} in a threshold form: 
\begin{align*}
x_i(\bt) &= \ind[\bigg]{v_i \geq \phi_{v_i}^{-1} \lrp[\Big]{\sum_{j \in N \sm
i} \phi_{\eta_{j \gets i}}(\eta_{j \gets i})}}
= \ind[\big]{v_i \geq \tauii(\bt_{-ii})} \\
    x_j(\bt) &= \ind[\bigg]{\eta_{i \gets j} \leq \phi_{\eta_{i \gets j}}^{-1} \lrp[\Big]{\phi_{v_j}(v_j) - \sum_{\mathclap{k\in N \sm \set{i,j}}} \phi_{\eta_{k\gets j}}(\eta_{k \gets j})}}
= \ind[\big]{\eta_{i \gets j} \leq \tauij(\bt_{-ij})}
\end{align*}
where we define the thresholds
\begin{displaymath}
\tauii(\bt_{-ii}) \eqdef \phi_{v_i}^{-1}\of[\Big]{\sum_{j \in N \sm i} \phi_{\eta_{j \gets i}}(\eta_{j\gets i})}
\hspace{0.8em}\text{and}\hspace{0.8em}
\tauij(\bt_{-ij}) \eqdef \phi_{\eta_{i \gets j}}^{-1}\of[\Big]{\phi_{v_j}(v_j) -\sum_{\mathclap{k\in N \sm \set{i,j}}} \phi_{\eta_{k \gets j}}(\eta_{k \gets j})}.
\end{displaymath}

Plugging these expressions into \eqref{eq:non-interim-payment1}, we have 
\begin{align*}
p_i(\bt)&= \ip[\big]{\bx(\bt)}{t_i} - \int_{\ubar{v}_i}^{v_i} x_i(\tilde{v}_i, \be_{\gets i})d\tilde{v}_i \notag \\
 &\quad + \sum_{j \in N \sm i} \int_{\ubar\eta_{i\gets j}}^{\eta_{i \gets j}} x_j(\tilde{\eta}_{i \gets j}, v_j, \set{\eta_{k \gets j}}_{k \neq i,j}) d\tilde{\eta}_{i \gets j} - C_i \notag\\
 &= \ip[\big]{\bx(\bt)}{t_i} - \int_{\ubar{v}_i}^{v_i} \ind*{\tilde{v}_i \geq \tauii(\bt_{-ii})}d\tilde{v}_i + \\
&\quad \quad \sum_{j \in N \sm i} \int_{\ubar\eta_{i\gets j}}^{\eta_{i \gets j}} \ind*{\tilde \eta_{i \gets j} \leq \tauij(\bt_{-ij})} d\tilde{\eta}_{i \gets j} - C_i. 
\end{align*}

Evaluating the integrals, plugging in the optimal constant $C_i$ and simplifying, we get 
\begin{align*}
    p_i(\bt) &= x_i(\bt) v_i - \sum_{j \in N \sm i} \eta_{i \gets j} x_j(\bt) - (v_i - \tauii(\bt)) x_i(\bt)\\
    & \quad \quad + \sum_{j \in N \sm i} \left[ \lrp*{\tauij(\bt_{-ij}) - \ubar{\eta}_{i \gets j}}(1-x_j(\bt)) + \lrp*{\eta_{i\gets j}-\ubar{\eta}_{i \gets j}} x_j(\bt) \right] - \sum_{j \in N\sm i}\ubar{\eta}_{i \gets j}\\
    &= x_i(\bt) \cdot \tauii(\bt_{-ii}) + \sum_{j \in N \sm i} \big(1-x_j(\bt)\big) \cdot \tauij(\bt_{-ij}).
\end{align*}

\subsection{Proof of Proposition \ref{prop:single_dim_rev_scen1}} \label{appendix:prop:single_dim_rev_scen1}

Under the assumptions of \cref{prop:single_dim_rev_scen1}, given allocation vector $\bx$ and payment $p_i$, the net utility of bidder $i$ with type $t_i$ simplifies to
\begin{equation}\label{eq:1d-factorization}
	v_i\cdot x_i - \sum_{j\neq i}\eta_{i\gets j}x_j - p_i
	=v_i\of[\bigg]{x_i-\alpha\sum_{j\neq i}x_j}-p_i.
\end{equation}
Furthermore, we see that the interim allocation of bidder $i$, $\by^{(i)}(t_i) \eqdef \cex{\bx(\bt)}{t_i} = \cex{\bx(\bt)}{v_i}$ is a function of $v_i$ only.
Similarly, we write $q_i(v_i)=\cex{p_i(\bt)}{t_i} = \cex{p_i(\bt)}{v_i}$ for the interim payment. Using~\eqref{eq:1d-factorization}, bidder $i$'s interim utility becomes
\begin{align*}
	\cex{u_i\big(\bx(\bt),\bp(\bt);\bt\big)}{t_i}
	&=v_i\cdot\ex[\Big]{x_i(\bt)-\alpha\sum\nolimits_{j\neq i}x_j(\bt)\given v_i}-q_i(v_i)\\
	&=v_i\cdot a_i(v_i) - q_i(v_i),
\end{align*}
where we defined the interim aggregated allocation effect
\begin{displaymath}
a_i(v_i) = \ex[\Big]{x_i(\bt)-\alpha\sum\nolimits_{j\neq i}x_j(\bt)\given v_i}.
\end{displaymath}
In other words, from the perspective of bidder $i$, the situation is exactly equivalent to the one of a single-dimensional allocation with linear utility.
The IC characterization of \Cref{prop:JMS99BNICchar} thus simplifies significantly and becomes essentially equivalent to the one of \cite{M81} (see also \citep[Prop.~3.6]{bonatti2022selling}).

\begin{lemma}\label{lemma:ic-1d}
	Assume that, for each $i\in N$, $t_i=v_i e_i - \sum_{j \in N \sm i} \eta_{i \gets j} e_j$ with $\eta_{i\gets j} =\alpha v_i$ for $j\in  N\sm i$. Then, the mechanism $(\bx, \bp)$ is BNIC iff for each bidder $i\in N$:
\begin{enumerate}[label=(\roman*)]
	\item the interim aggregated allocation effect $a_i$ is non-decreasing.
	\item the interim payment is given by
		\begin{equation}\label{eq:int-payment}
			q_i(v_i) = v_i\cdot a_i(v_i)-\int_{\ubar{v}_i}^{v_i} a_i(s)ds+C_i,
		\end{equation}
		for some constant $C_i \in \R$.
\end{enumerate}
\end{lemma}

\begin{proof}[Proof of \cref{prop:single_dim_rev_scen1}]
Using \eqref{eq:int-payment}, we first compute the expected payment of firm $i$
in terms of the virtual value function $\phi_i$ associated with $v_i$'s
distribution (cf.~\cite[Lemma~4.4]{bonatti2022selling}):
\begin{align*}
\ex{p_i(\bt)}
&= \ex[\big]{\ex{p_i(\bt)\given v_i}} = \ex{q_i(v_i)}\\
&=\int_{\ubar{v}_i}^{\bar v_i}v_i a_i(v_i) f_i(v_i)dv_i
	-\int_{\ubar{v}_i}^{\bar v_i}\int_{\ubar{v}_i}^{v_i} a_i(s)f_i(v_i)dsdv_i+C_i\\
&=\int_{\ubar{v}_i}^{\bar v_i}v_i a_i(v_i) f_i(v_i)dv_i
	-\int_{\ubar{v}_i}^{\bar v_i}a_i(s)\left(\int_{s}^{\bar v_i}f_i(v_i)dv_i\right)ds+C_i\\
&=\int_{\ubar{v}_i}^{\bar v_i}v_i a_i(v_i) f_i(v_i)dv_i
	-\int_{\ubar{v}_i}^{\bar v_i}a_i(s)\big(1-F_i(s)\big)ds+C_i\\
&=\int_{\ubar{v}_i}^{\bar
v_i}a_i(v_i)\left(v_i-\frac{1-F_i(v_i)}{f_i(v_i)}\right) f_i(v_i)dv_i+C_i,
\end{align*}
where we swapped the order of integration in the third line. We recognize the virtual value function $\phi_i$ in the last line, hence
\begin{displaymath}
	\ex{p_i(\bt)} =\E[a_i(v_i)\phi_i(v_i)]+C_i.
\end{displaymath}

Note that using the law of total expectation
\begin{displaymath}
\ex{a_i(v_i)\phi_i(v_i)} =
\ex*{\phi_i(v_i)\left(x_i(\bt)-\alpha\sum\nolimits_{j\neq i}x_j(\bt)\right)},
\end{displaymath}
which implies the following expression for the expected revenue
\begin{align*}
	&\ex*{\sum_{i\in N}\phi_i(v_i)\left(x_i(\bt)-\alpha\sum\nolimits_{j\neq i}x_j(\bt)\right)}
	+\sum_{i\in N}C_i\\
	&=\ex*{\sum_{i\in N}x_i(\bt)\left(\phi_i(v_i)-\alpha\sum\nolimits_{j\neq i}\phi_j(v_j)\right)}
	+\sum_{i\in N}C_i
\end{align*}
where we swapped the order of summation on the right-hand side. This
immediately implies that the optimal allocation is given by
\begin{displaymath}
	x_i(\bt)= \ind*{\phi_i(v_i)\geq \alpha\sum\nolimits_{j\neq i}\phi_j(v_j)}.
\end{displaymath}

We conclude by verifying that the resulting mechanism is BNIC. For this, it
suffices to check that the aggregated allocation effect is non-decreasing.
Denoting by $G_i$ the c.d.f.\ of $\sum_{j\in N\sm i}\phi_j(v_j)$ and by
$G_{i,j}$ the c.d.f\ of $\sum_{k\in N\sm\set{i,j}} \phi_k(v_k)$, we compute
(cf.~\cite[Prop~4.6]{bonatti2022selling})
\begin{displaymath}
	a_i(v_i)= G_i\big(\phi_i(v_i)/\alpha\big)
	-\alpha\sum_{j\in N\sm i}\ex*{G_{i,j}\big(\phi_j(v_j)/\alpha-\phi_i(v_i)\big)\given v_i}.
\end{displaymath}
Since cumulative distribution functions are non-decreasing and by regularity of the distribution of $v_i$, the function $v_i\mapsto G_i(\phi_i(v_i)/\alpha)$ is non-decreasing and $v_i\mapsto G_{i,j}\big(\phi_j(v_j)/\alpha-\phi_i(v_i)\big)$ is non-increasing (for each $v_j$), which concludes the proof.
\end{proof}

\subsection{Proof of Theorem~\ref{prop:rev}} \label{appendix:rm2_proof}

	We consider a mechanism $(\bx$, $\bp)$ and use the BNIC characterization of
	Proposition~\ref{prop:BNIC_char}. In particular, recall that there exists
	a non-decreasing function $\ty_i$ such that $y_i(\bt)=\ty_i(v_i)$, where
	$y_i$ is the interim allocation. Plugging in the form of interim payments
	$q_i$ given by \eqref{eq:prop3_payment_app} we get
\begin{equation}
	\label{eq:foo}
	\rev(\bx, \bp)=\sum_{i\in N}
	\E\left[v_i \ty_i(v_i)
		- \int_{\ubar v_i}^{v_i} \ty_i(v)dv
	-\sum_{j\in N\sm i}\E\left[\eta_{i\gets j}\cdot x_j(\bt)\,\middle|\,t_i\right]
	- C_i\right]
	\,.
\end{equation}
Observe that the last term on the right-hand side is independent of the choice
of $(\bx,\bp)$ and can thus be ignored when searching for the revenue optimal
auction.

For the second term, swapping the order of integration gives
\begin{align*}
	\E\left[\int_{\ubar v_i}^{v_i} \ty_i(v)dv \right]
	&=\int_{\ubar v_i}^{\bar v_i} f_{v_i}(v_i)\left(\int_{\ubar v_i}^{v_i} \ty_i(v)dv\right)dv_i\\
	&=\int_{\ubar v_i}^{\bar v_i} \big(1-F_{v_i}(v_i)\big)\ty_i(v_i)dv_i
	=\E\left[\frac{1-F_{v_i}(v_i)}{f_{v_i}(v_i)}\cdot\ty_i(v_i)\right]
	\,.
\end{align*}

For the third term, we write
\begin{align*}
	&\E\left[\sum_{i\in N}\sum_{j\in N\sm i}
	\E\left[\eta_{i\gets j}x_j(\bt)\,\middle|\,t_i\right]\right]
	= \sum_{i\in N}\sum_{j\in N\sm i}
	\E\big[\eta_{i\gets j}\cdot x_j(\bt)\big]\\
	  &\quad\quad=\sum_{i\in N}\sum_{j\in N\sm i}\E\big[\eta_{j\gets i}\cdot x_i(\bt)\big]
	=\sum_{i\in N}\sum_{j\in N\sm i}\E\big[\eta_{j\gets i}\cdot
	\E[x_i(\bt)\,|\,t_i]\big]\\
	&\quad\quad=\sum_{i\in N}\sum_{j\in N\sm i}\E\big[\eta_{j\gets i}\cdot
	\ty_i(v_i)\big]
	=\sum_{i\in N}\sum_{j\in N\sm i}\E\big[\ty_i(v_i)\cdot\E[\eta_{j\gets
	i}\,|\,v_i]\big]\,,
\end{align*}
where the first, third and last equality use the law of total expectation, the
second equality is just a change of index and the penultimate is by definition
of $\ty_i$.

Combining the previous derivations, we get that the revenue maximizing problem
is equivalent to maximizing
\begin{equation}
	\label{eq:opt}
	\sum_{i\in N}
	\E\left[\ty_i(v_i)\left(v_i - \frac{1-F_{v_i}(v_i)}{f_{v_i}(v_i)}
	-\sum_{j\in N\sm i}\E[\eta_{j\gets i}\,|\, v_i]\right)\right]
	\,.
\end{equation}
where $\ty_i$ is the interim allocation  computed from $x_i$ and must be
non-decreasing by Proposition \ref{prop:BNIC_char}. Hence, we see that the objective function as well as the
BNIC and IR constraints can be written solely in terms of the functions
$(\ty_i)_{i\in N}$. It is thus sufficient to optimize over each summand separately,
under the constraint that $\ty_i$ be non-decreasing and $[0,1]$-valued.

Under the regularity condition that $\widetilde\phi_i: v_i\mapsto\phi_i(v_i) -\sum_{j\in N\sm i}\E[\eta_{j\gets i}\,|\, v_i]$ is non-decreasing, observe that each summand in \eqref{eq:opt} can be written concisely in terms of $\widetilde\phi_i(v_i)$ as $\sup_{y} \E[\widetilde\phi_i(v_i)\tilde y_i(v_i)]$. The choice $\tilde y_i(v_i)= \ind{\widetilde\phi_i(v_i)\geq 0}$ maximizes the integrand pointwise and is non-decreasing in $v_i$ since $\widetilde\phi_i$ is also non-decreasing. Hence it also maximizes the expectation subject to the monotonicity constraint, yielding the optimal BNIC allocation rule.

To complete the proof we need to choose the smallest constant of integration $C_i$ in \eqref{eq:foo} such that interim IR is satisfied.  By Proposition~\ref{prop:IRchar}, it suffices to set $C_i$ to be the lowest interim utility a bidder could get in any outside option, which is exactly $-\sum_{j\in N\sm i} \E[\eta_{i \gets j}]$.

\end{document}